\documentclass[fleqn,usenatbib]{mnras}

\usepackage{newtxtext,newtxmath}

\usepackage[T1]{fontenc}
\usepackage{ae,aecompl}

\usepackage{color}

\usepackage{graphicx}
\usepackage{aas_macros}
\usepackage{amsmath}
\usepackage{amssymb}
\usepackage[normalem]{ulem}
\usepackage{tabularx}
\usepackage{times}

\hypersetup{final}

\newcommand{\eagle}{\textsc{eagle}}
\newcommand{\auriga}{\textsc{auriga}}
\newcommand{\MW}{\textsc{MW}}

\newcommand{\smallspace}{\vspace{-.35cm}}

\newcommand{\kpc}{\ensuremath{~\text{kpc}}}

\newcommand{\Msun}{\ensuremath{~\text{M}_\odot}}
\newcommand{\kms}{\ensuremath{~\text{km~s}^{-1}}}

\newcommand{\FEL}{\ensuremath{F(\widetilde{E},\widetilde{L})}}
\newcommand{\FEE}{\ensuremath{F_E(\widetilde{E})}}
\newcommand{\FLL}{\ensuremath{F_L(\widetilde{L})}}
\newcommand{\EL}{\ensuremath{(\widetilde{E},\widetilde{L})}}
\newcommand{\Etilde}{\ensuremath{\widetilde{E}}}
\newcommand{\Ltilde}{\ensuremath{\widetilde{L}}}

\newcommand{\change}[1]{{#1}}
\title[Mass of the MW from satellites]{The mass of the Milky Way from satellite dynamics}

\author[T. M. Callingham et al.]
{\parbox{\textwidth}{
Thomas M. Callingham,$^{1}$\thanks{E-mail: thomas.m.callingham@durham.ac.uk}
Marius Cautun,$^{1}$
Alis J. Deason,$^{1}$
Carlos S. Frenk,$^{1}$
Wenting Wang,$^{2}$
Facundo A. G\'{o}mez,$^{3,4}$
Robert J. J. Grand,$^{5,6}$
Federico Marinacci$^{7}$
and 
Ruediger Pakmor$^{5}$
\vspace{.20cm}} \\
$^{1}$Institute of Computational Cosmology, Department of Physics, University of Durham, South Road, Durham DH1 3LE, UK \\
$^2$Kavli IPMU (WPI), UTIAS, The University of Tokyo, Kashiwa, Chiba 277-8583, Japan \\
$^{3}$Instituto de Investigaci{\'o}n Multidisciplinar en Ciencia y Tecnolog{\'i}a, Universidad de La Serena, Ra{\'u}l Bitr{\'a}n 1305, La Serena, Chile \\
$^{4}$Departamento de F{\'i}sica y Astronom{\'i}a, Universidad de La Serena, Av. Juan Cisternas 1200 N, La Serena, Chile \\
$^5$Heidelberger Institut f\"{u}r Theoretische Studien,
Schloss-Wolfsbrunnenweg 35, 69118 Heidelberg, Germany \\
$^6$Zentrum f\"{u}r Astronomie der Universit\"{a}t Heidelberg, ARI,
M\"{o}nchhofstr. 12-14, 69120 Heidelberg, Germany \\
$^7$Department of Physics, Kavli Institute for Astrophysics and Space Research, MIT, Cambridge, MA 02139, USA \\
}

\pubyear{2018}

\begin{document}
\label{firstpage}
\pagerange{\pageref{firstpage}--\pageref{lastpage}}
\maketitle

\begin{abstract}
  We present and apply a method to infer the mass of the Milky Way
  (MW) by comparing the dynamics of MW satellites to those of model
  satellites in the \eagle{} cosmological hydrodynamics simulations. A
  distribution function (DF) for galactic satellites is constructed
  from \eagle{} using specific angular momentum and specific energy,
  which are scaled so as to be independent of host halo mass. In this
  2-dimensional space, the orbital properties of satellite galaxies
  vary according to the host halo mass. The halo mass can be inferred
  by calculating the likelihood that the observed satellite population
  is drawn from this DF. Our method is robustly calibrated on mock
  \eagle{} systems. We validate it by applying it to the completely
  independent suite of 30 \auriga{} high-resolution simulations of
  MW-like galaxies: the method accurately recovers their true mass and
  associated uncertainties. We then apply it to ten classical
  satellites of the MW with 6D phase-space measurements, including
  updated proper motions from the \textit{Gaia} satellite. The mass of
  the MW is estimated to be
  $M_{200}^{\textnormal{MW}}=1.17_{-0.15}^{+0.21}\times10^{12}M_{\odot}$ (68\% confidence limits). \change{We
  combine our total mass estimate with recent mass estimates in the
 inner regions of the Galaxy to infer an inner dark matter (DM) mass fraction
 $M^\textnormal{DM}(<20~\rm{kpc})/M^\textnormal{DM}_{200}=0.12$
  which is typical of ${\sim}10^{12}M_{\odot}$ $\Lambda$CDM haloes in hydrodynamical galaxy formation simulations. Assuming an NFW profile, this is equivalent to a halo concentration of
  $c_{200}^{\textnormal{MW}}=10.9^{+2.6}_{-2.0}$.}
\end{abstract}



\section{Introduction}
The mass of the Milky Way (MW) is a fundamental astrophysical
parameter. It is not only important for placing the MW in context
within the general galaxy population, but it also plays a major role
when trying to address some of the biggest mysteries of modern
astrophysics and cosmology. The intricacies of galaxy formation are
highly dependant on feedback and star formation processes, which
undergo a crucial physical transition around the MW mass
\citep[e.g.][]{Bower2017}. Apparent discrepancies with the standard
$\Lambda$CDM model, such as the missing satellites
\citep{Klypin1999,Moore1999} and the too-big-to-fail problems
\citep{BK11_TBtF_MNRAS.415L..40B} depend strongly on the MW halo mass
\citep[e.g.][]{Purcell2012,Wang12_MissingSat_MNRAS.424.2715W,VeraCiro2013,
  Cautun2014b}. In addition, tests of alternative warm 
  dark matter models
\citep{Kennedy2014:WDMParticlefromMWSat_MNRAS.442.2487K,Lovell2014:WarmDM_MNRAS.439..300L}
are also subject to the total halo mass. Thus, by considering the
cosmological context of the MW and its population of dwarf galaxy
satellites, important inferences about large-scale cosmology can be
made. With the recent \textit{Gaia} DR2 release
\citep{GaiaDR22018arXiv180409365G}, we now have significantly more
information than ever before about our galaxy, and are better placed
to make progress on these problems.


There have been many attempts to infer directly the MW
mass through a variety of methods. The total MW mass is dominated by
its dark matter (DM) halo, which cannot be observed directly. Instead,
its properties must be inferred from the properties of luminous
populations, such as the luminosity function of MW satellites
\citep[mostly the Large and Small Magellanic Clouds,
e.g.][]{Busha2011b,Gonzalez2013,Cautun2014MNRAS.445.2049C} and the kinematics of various
dynamical tracers of the Galactic halo. 
The dynamics of halo tracers are mostly determined by the
gravitational potential of the MW halo, and provide a key indirect
probe of the total halo mass. Examples of halo tracers used for this
purpose are satellite galaxies \citep[e.g.][]{WilkinsonEvans1999,
  Watkins2010}, globular clusters \citep[e.g.][]{Eadie2016, BinneyWong2017,
  Sohn2018, WatkinsMWGC2018arXiv180411348W}, halo stars
\citep[e.g.][]{Xue2008,Deason12_BrokenDegenMNRAS.424L..44D, Kafle2012,Kafle2014}, high velocity stars
\citep[e.g.][]{Smith2007,
  Piffl2014,Fragione17_HVSMWMassNewA...55...32F,Rossi2017,Monari2018} and stellar streams
\citep[e.g.][]{Koposov2010, Newberg10_StreamMWMass_ApJ...711...32N, Gibbons2014,
  Kupper2015, Bowden2015}.

There are a variety of methods for inferring the Galactic halo mass
using dynamical tracers. A common approach is to model the tracers as
distributions in equilibrium whose parameters are determined by
fitting the model to observational data \citep[e.g.][]{Evans2003,Han2016a}.
Advances in the calculation of action-angle coordinates
\citep[e.g.][]{Vasiliev18_AGAMAarXiv180208239V} have led to a new
generation of analytical galaxy modelling, centred around distribution
functions (DFs) in action-angle space.  Examples include modelling the
MW population of globular clusters
\citep[e.g.][]{PostiHelmi2018arXiv180501408P} or individual DFs of
components such as the thick and thin disc, bulge, stellar halo and DM
halo \citep{ColeBinneyCore2017MNRAS.465..798C}.
The recent availability of large cosmological simulation has enabled a new class
of methods based on comparing the observed properties of MW satellites
to those of substructures in cosmological simulations
\citep[e.g.][]{Busha2011ApJ...743...40B,Patel2017Orbits2_MNRAS.468.3428P}.

Although over the past decades a large amount of effort has been
dedicated to inferring the Galactic halo mass, its value remains
uncertain to within a factor of two, with most mass estimates ranging
from $0.5$ to $2.5\times10^{12}M_{\odot}$ \citep[e.g.][and our
Fig. \ref{fig:LitComparison}]{Wang15_DMHaloDynTracers_MNRAS.453..377W}. While
many studies claim uncertainties smaller than this range,
the analytical models upon which they rely require several assumptions
such as dynamical equilibrium and a given shape of the density
or the velocity anisotropy profiles. These assumptions can lead to
additional systematic errors, which  are difficult to quantify but can be the dominant
source of error \citep[e.g. see][]{Yencho2006, Wang15_DMHaloDynTracers_MNRAS.453..377W,Wang2018}.
\change{This is especially true for the MW halo whose dynamics are likely to be  affected by the presence of a very massive satellite, the Large Magellanic Cloud \citep{Gomez2015,Penarrubia2016,Shao2018b}.} 
Furthermore, most methods
typically estimate the mass within the inner tens of kiloparsecs, since this
is the region where most tracers (such as halo stars and globular
clusters) reside, necessitating an extrapolation to the virial
radius. This extrapolation requires additional
assumptions about the radial density profile of the MW and can lead to
further systematic uncertainties.

Large-volume high-resolution cosmological simulations offer a unique
test-bed for analytical mass determination methods \citep[e.g.][]{Han2016b,Penarrubia2017,Wang2017} and, importantly,
enable new methods for inferring the Galactic halo mass with a minimal
set of assumptions. The simulations have the advantage of
self-consistently capturing the complexities of halo and galaxy
formation, as well as the effects of halo-to-halo variation. However,
with a few exceptions, the limited mass resolution of current
simulations means that they can resolve satellite galaxies but not
halo stars or globular clusters \citep[although see
e.g.][]{Pfeffer2018,Grand2018}. This is not a major limitation since
satellite galaxies, due to their radially extended spatial distribution, are one of the best probes of the outer MW
halo. This is especially true now that the \emph{Gaia} DR2 release has
provided a large sample of MW satellites with full 6D phase
space information
\citep[][]{GaiaDR2_MotionsDwarfGC_2018arXiv180409381G,
  Fritz2018,Simon2018}.

Galactic halo mass estimates that rely on cosmological simulations are
relatively recent. \citet{Busha2011ApJ...743...40B} pioneered the
approach of inferring halo properties by finding the best match
between the MW satellites and satellites of simulated haloes. The MW
mass is then determined by weighting the host haloes according to the
quality of the satellite match, a technique known as importance
sampling. \citeauthor{Busha2011ApJ...743...40B} used the distance,
velocity and size of the Large and Small Magellanic Clouds (hereafter LMC and SMC) to
constrain the MW mass. The distance and velocity of satellites can
vary rapidly, especially when close to the pericentre of their orbit,
so very large simulations are needed in order to find enough
counterparts to the MW system.

\citet{Patel2017Orbits2_MNRAS.468.3428P} pointed out that
approximately conserved quantities, such as angular momentum, are
better for identifying satellite analogues in simulations. This makes
it easier to find MW counterparts; applying the criterion to a larger
number of satellites results in a more precise mass determination
\citep{PatelMWEnsemble2018ApJ...857...78P}. A further advance was
achieved by \citet{Li2017ApJ...850..116L} who showed that, when scaled
appropriately, the DF of satellite energy and angular momentum becomes
independent of halo mass. This scaling allows for a more efficient use
of simulation data, since any halo can be rescaled to a different
mass, and thus a better sampling of halo formation histories and
halo-to-halo variation can be achieved. This approach represents a
major improvement over importance sampling methods, in which the
statistically relevant systems are those in a small mass range.

In this paper we improve and extend the \citet{Li2017ApJ...850..116L}
mass determination method. We start by constructing the phase-space
distribution of satellite galaxies using a very large sample of host
haloes taken from the \eagle{} (Evolution and Assembly of
GaLaxies and their Environments) galaxy formation simulation
\citep{EAGLE2015MNRAS.446..521S,Crain2015}. We then describe and calibrate three
mass inference methods based on the satellite distributions of: i)
angular momentum only, ii) energy only, and iii) a combination of both
angular momentum and energy. We test these methods by applying them to
an independent set of simulations, taken from the \auriga{} project
\citep{Auriga_2017MNRAS.467..179G}; this is a very stringent test
because of the much higher resolution and rather different galaxy
formation model implemented in \auriga{} compared to
\eagle{}. Finally, we apply our methods to the latest observations of
the classical satellites to determine the MW halo mass; we are able to
estimate this mass with an uncertainty of only $20\%$.

The structure of the paper is as follows. Section
\ref{sec:satellite_phase_space} describes the construction of the
phase-space DFs using the \eagle{} data. Section~\ref{sec:method}
describes our mass inference methods, their calibration and validation
with tests on mock systems. In Section~\ref{sec:MW}, we apply this
method to the observed MW system and discuss our results. Finally,
Section~\ref{sec:conclusion} summarises and concludes the paper.

\section{Construction of the satellite distribution}
\label{sec:satellite_phase_space}

We now describe how to obtain a phase space distribution of
satellites that, when scaled appropriately, is independent of host
halo mass. We then introduce the MW observations, and the simulation
data that we use for calculating the phase-space distribution function
of satellite galaxies. 

\subsection{Theoretical background}
We are interested in the energy and angular momentum distribution of
Galactic satellites. This can be calculated starting from the observed
distance, $r^{s}$, tangential velocity, $v_{t}^{s}$, and speed,
$v^{s}$, of satellite $s$, which we use to define the vector: 
\begin{equation}
	\boldsymbol{x}^{s}=\left(v^{s},v_{\textnormal{t}}^{s},r^{s}\right).
\end{equation}
The specific energy, $E$, and specific angular momentum, $L$, of a satellite are given by:
\begin{equation} \label{eq:EnergyEq}
	\begin{aligned}
	E &= \frac{1}{2}\left|\textnormal{\textbf{v}}\right|^{2}+\Phi\left(r\right)\\
	L &= \left|\boldsymbol{r}\times\boldsymbol{v}\right|=rv_{\textnormal{t}},
	\end{aligned}
\end{equation}
where $\Phi(r)$ is the gravitational potential at the position of the satellite. This cannot be measured directly in observations, and to calculate it we need to assume a mass profile for the host halo. 
Here, we assume that the host density profile is well approximated by
a spherically symmetric Navarro, Frenk and White profile (hereafter
NFW; \citealt{Navarro1996,Navarro1997}), whose gravitational potential
is given by: 
\begin{equation} \label{eq:NFWPotEq}
	\Phi_{\textnormal{NFW}}\left(r\right)=-\frac{GM_{200}
    }{r}\frac{\ln\left(1+C\frac{r}{R_{200}}\right)}{\ln\left(1+C\right)-\frac{C}{C+1}}
    \;,
\end{equation}
where $C$ is the concentration of the
halo and $M_{200}$ and $R_{200}$ denotes the halo mass and radius, respectively. The mass, $M_{200}$,
corresponds to the mass enclosed within a sphere of average density
$200$ times the critical density. 

The NFW profile provides a good description of the radial density
profile of relaxed haloes in DM-only simulations. The addition of
baryons leads to a contraction of the inner region of haloes, and thus
to a systematic departure from an NFW profile
\citep[e.g.][]{Gnedin2004}. However, at large enough distances
(e.g. $r\gtrsim 20\kpc$ for a halo mass of $10^{12}\Msun$) the NFW profile
still provides a very good description of the mass distribution even
in galaxy formation simulations
\citep[e.g.][]{MatthieuNFWConc2015MNRAS.451.1247S,Zhu2016}. In this work, we
consider only satellites relatively far from the halo centre, where
the NFW function represents a good approximation of the mass profile.
 
DM haloes have several self-similar properties, such as their density
profiles \citep[e.g.][]{Navarro1996,Navarro1997}, the substructure
mass function
\citep[e.g.][]{Wang12_MissingSat_MNRAS.424.2715W,Cautun2014b} and the
radial number density of subhaloes
\citep{Springel2008,Hellwing2016}. \citet{Li2017ApJ...850..116L}
showed that the same self-similar behaviour also holds for the energy
and angular momentum distribution functions of subhaloes. This implies
that, when scaled accordingly, satellites around hosts of different
mass follow the same energy and angular momentum distribution. The
same self-similar behaviour also holds to a good approximation in the
\eagle{} hydrodynamic simulation (see Appendix~\ref{Appendix:Scaling}). 

For a self-similar halo density profile, the satellites' positions and
velocities scale with $M_{200}^{1/3}$ \citep{Li2017ApJ...850..116L}. A
given host halo and its associated satellite system, can therefore be
scaled to a different host halo mass, $M_{200}^{\textnormal{Scale}}$,
as: 
\begin{equation} \label{eq:scaleddisv}
    \left(r',v',v_{\textnormal{t}}'\right)=\left(\frac{M_{200}^{\textnormal{Scale}}}{M_{200}}\right)^{\frac{1}{3}}\left(r,v,v_{\textnormal{t}}\right)
    \;.
\end{equation}

This implies that the energy and angular momentum of satellites also
scale with halo mass through the relation $E,L\propto
M_{200}^{2/3}$. Thus, we can choose characteristic $E_0$ and $L_0$
values for each halo mass and use them to rescale the $E$ and $L$
values of each satellite to obtain mass independent quantities. For
each halo, we define the scaled specific energy, $\widetilde{E}$, and
scaled specific angular momentum, $\widetilde{L}$, as: 
\begin{equation} \label{eq:sEsLEq}
	\left(\widetilde{E},\widetilde{L}\right) = \left(\frac{E}{E_0},\frac{L}{L_0}\right)
    \;,
\end{equation}
where the characteristic $E_0$ and $L_0$ values correspond to the
energy and angular momentum of a circular orbit at $R_{200}$ and are
given by: 
\begin{equation}
\begin{aligned}
	E_0 &= \frac{GM_{200}}{R_{200}}\\
	L_0 &= \sqrt{GM_{200} R_{200}}.
\end{aligned}
\end{equation}
This scaling relation preserves the relaxation state, concentration
and formation history of the halo, giving scaled properties that are
independent of host mass (see Appendix~\ref{Appendix:Scaling}).

\subsection{Observational data for the MW satellites}
\label{Sec:ObsData}

We aim to estimate the MW halo mass using the classical satellites
since those have the best proper motion measurements. The method we
employ is flexible enough to incorporate the ultrafaint dwarfs;
however, the \eagle{} simulation, which we use for calibration, does
not resolve the ultrafaint satellites.  Furthermore, we discard any
satellites closer than $40\kpc$ (see section \ref{Sec:EagleSims}), so
we exclude the Sagittarius dwarf from our observational
sample. Sagittarius is currently at a distance of $26\kpc$, undergoing
strong tidal disruption by the MW disc, and is therefore unsuitable 
as a tracer of the DM halo. This leaves 10 classical satellites with
adequate kinematical data (see Table \ref{Tab:SatObs}).

\begin{table}
	\centering
	\caption{\change{Properties of the  classical Galactic satellites
        used in this work. The last two columns give the calculated energy and angular momentum values for each satellite.} The energy has been calculated using an
        NFW profile with a concentration of 8, for a mass,
        $M_{200}^{\textnormal{MW}}=1.17\times10^{12}M_{\odot}$,
        which corresponds to our best MW-halo mass estimate. The
          distance is with respect to the Galactic Centre. The
          specific orbital angular momentum, $L$, and specific energy,
          $E$, of the satellites are expressed in terms of the angular
          momentum, $L_{0;\;\textnormal{MW}}$, and energy,
          $E_{0;\;\textnormal{MW}}$, of a circular orbit at the virial
          radius, $R_{200}$. For the mass and concentration assumed
          here, we have   $L_{0;\;\textnormal{MW}} = 3.34\times10^{4} 
    \kpc\kms$  and $E_{0;\;\textnormal{MW}} =
    2.28\times10^{4}\textnormal{km}^{2}~\textnormal{s}^{-2}$. The
    errors give the $68\%$ confidence interval based on Monte Carlo
    sampling of the observational errors (see text for details). 
	}
	\label{Tab:SatObs}
    \setlength\extrarowheight{5pt}
\begin{tabularx}{0.83\columnwidth}{l c r r}

\hline
Satellite & Distance $[\rm{kpc}]$ & $L/L_{0;\;\textnormal{MW}}$ & $E/E_{0;\;\textnormal{MW}}$\tabularnewline[.1cm]
\hline
LMC & $51\pm 2$ & $0.46_{-0.05}^{+0.05}$ & $-1.33_{-0.31}^{+0.32}$\tabularnewline
SMC & $64\pm 4 $ & $0.46_{-0.08}^{+0.08}$ & $-1.84_{-0.37}^{+0.42}$\tabularnewline
Draco & $76\pm 6$ & $0.30_{-0.03}^{+0.03}$ & $-2.40_{-0.11}^{+0.10}$\tabularnewline
Ursa Minor & $76\pm 6$ & $0.32_{-0.01}^{+0.02}$ & $-2.39_{-0.05}^{+0.05}$\tabularnewline
Sculptor & $86\pm 6$ & $0.48_{-0.03}^{+0.03}$ & $-1.89_{-0.07}^{+0.07}$\tabularnewline
Sextans & $86\pm 4$ & $0.67_{-0.05}^{+0.06}$ & $-1.21_{-0.16}^{+0.17}$\tabularnewline
Carina & $105\pm 6\;$ & $0.55_{-0.08}^{+0.08}$ & $-1.86_{-0.19}^{+0.19}$\tabularnewline
Fornax & $147 \pm 12$ & $0.70_{-0.19}^{+0.21}$ & $-1.52_{-0.30}^{+0.33}$\tabularnewline
Leo II & $233 \pm 14$ & $0.96_{-0.28}^{+0.30}$ & $-1.20_{-0.21}^{+0.29}$\tabularnewline
Leo I & $254 \pm15 $ & $0.82_{-0.26}^{+0.28}$ & $-0.67_{-0.15}^{+0.21}$\tabularnewline[0.05cm]
\hline
\end{tabularx}
\end{table}

We take satellite positions, distances and radial velocities from the
\cite{Obs_McConnachie12_DwarfPosAJ....144....4M} compilation. We use
the observed proper motions of the classical satellites derived from
the \textit{Gaia} data release DR2
\citep{GaiaDR2_MotionsDwarfGC_2018arXiv180409381G}, apart from the
most distant satellites, Leo I and Leo II, for which we use the Hubble
Space Telescope proper motions \citep{Obs_PMLeoI_2013ApJ...768..139S,
  Obs_PMLeoII_2016AJ....152..166P} since these have smaller
uncertainties.  

To calculate the energy and angular momentum, we transform the
satellite positions and velocities from Heliocentric to Galactocentric
coordinates using the procedure described in \citet{Cautun2015}. The
transformation depends on the Sun's position and velocity for which we
adopt: $d = 8.29 \pm 0.16 \kpc$ for the distance of the Sun from the
Galactic Centre; $V_{\text{circ}} = 239 \pm 5 \kms$ for the circular
velocity at the Sun's position \citep{McMillan2011}; and
$(U,V,W) = (11.1 \pm 0.8, 12.2 \pm 0.5, 7.3 \pm 0.4) \kms$ for the
Sun's motion with respect to the local standard of rest,
\citep{Schonrich2010}. When transforming to Galactocentric coordinates
we account for errors in the distance, radial velocity and proper
motion of each satellite, as well as in the Sun's position and
velocity, which we model as normally distributed errors. To propagate
the errors, we generate a set of 1000 Monte Carlo realizations of the
MW system in heliocentric coordinates and transform each realization
to Galactocentric coordinates.

\subsection{EAGLE simulation sample}
\label{Sec:EagleSims} 

We select our sample of host haloes and satellite populations from the
reference run of the \eagle{} project \citep{EAGLE2015MNRAS.446..521S,Crain2015}. The simulation follows
galaxy formation in a $100\,\textnormal{Mpc}$ cubic volume with the
Planck cosmological parameters \citep[][ see Table 9]{planck2014}
using $1504^3$ dark matter particles of mass of
$9.7\times10^{6}M_{\odot}$ and $1504^3$ gas particles of initial mass
of $1.81\times10^{6}M_{\odot}$. \eagle{} models the relevant baryonic
physics processes such as gas cooling, stochastic star formation,
stellar and AGN feedback, and the injection of metals from supernovae
and stellar winds; it was calibrated to reproduce the present day
stellar mass function, galaxy sizes and the galaxy mass -- black hole
mass relation. The population of haloes and subhaloes was identified
using the \textsc{subfind} algorithm \citep{Springel2001}.  The large
volume of the \eagle{} simulation provides a large sample of haloes,
of a wide range of masses and assembly histories. Our final sample consists of the
following host haloes and satellites galaxies.

\subsubsection*{Selection criteria for hosts haloes:} 
\begin{enumerate}
	\item Halo mass, $M_{200}$,  in the range $10^{11.7}\Msun$ to
          $10^{12.5}\Msun$; 
	\item  relaxed systems, that is haloes for which  the distance
          between the centre of mass and the centre of potential 
          is less than $0.07R_{200}$ and the total mass in
          substructures is less than $10\%$
          \citep[][]{Neto2007}. 
\end{enumerate}

\subsubsection*{Selection criteria of satellite galaxies:} 
\begin{enumerate}
	\item Distance from halo centre in the range $40\kpc < r' < 300\kpc$,
  	where $r'= r (10^{12}\Msun/M_{200})^{1/3}$ is the rescaled distance
  	of the satellite corresponding to a halo of mass $10^{12}\Msun$ (see
  	equation \ref{eq:scaleddisv}); this results in a similar radial
  	distribution as the MW satellites if the MW halo had a mass of
  	$10^{12}\Msun$; 

	\item the satellite is luminous, i.e it contains at least one star particle, which excludes dark subhaloes.

\end{enumerate}
This gives a sample of approximately ${\sim}1,200$ host haloes and
${\sim}14,000$ satellites. Our mass scaling method allows us to choose
haloes in a broad mass range. The restriction on the radial
distribution of satellite galaxies is chosen so that the model samples
matches the observed one and to ensure that the potential is dominated
by DM. 

\begin{figure}
	\includegraphics[width=\columnwidth]{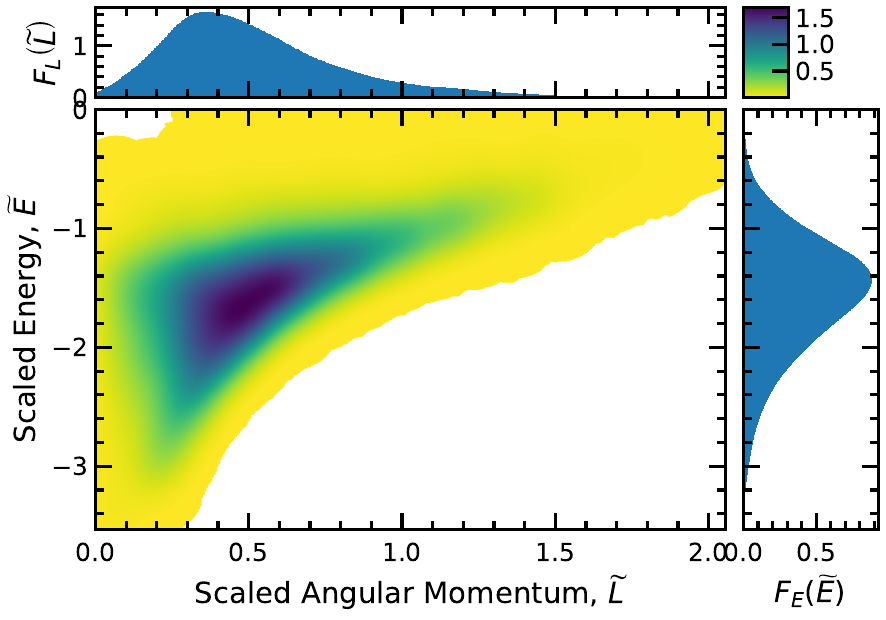}
    \smallspace{}
    \caption{ The distribution, \FEL{}, 
    of bound \eagle{} satellites in terms of the scaled angular
    momentum, \Ltilde{}, and scaled energy, \Etilde{}. The energy and
    angular momentum are scaled according to equation
    \eqref{eq:sEsLEq} to obtain quantities that are independent of
    host halo mass, $M_{200}$. The colour gives the number density of
    satellites, with dark colours corresponding to higher number
    densities (see colour bar). The two side panels show the one
    dimensional distributions of the scaled energy \FEE{}  
    (right-hand panel) and scaled angular momentum \FLL{} 
    (top panel) of satellites.  
    }
    \label{fig:ELDist}
\end{figure}

In Fig. \ref{fig:ELDist} we show the distribution of \eagle{}
satellites in scaled energy and angular momentum space,
($\widetilde{E}, \widetilde{L}$). 
For each satellite, we calculate the energy by assuming that the host halo is well described by an NFW profile 
\change{individually fitted to each halo as described in \citet{MatthieuNFWConc2015MNRAS.451.1247S}}. This procedure is similar to how energy is calculated for observational satellites, and thus allows for a proper comparison between theory and observations.
To obtain a continuous DF, we applied a 2D Gaussian smoothing with dispersions
$\alpha\sigma_{\widetilde{L}}$ and $\alpha\sigma_{\widetilde{E}}$ for
the $\widetilde{L}$ and $\widetilde{E}$ directions, respectively. The
symbols $\sigma_{\widetilde{L}}=0.36$ and
$\sigma_{\widetilde{E}}=0.52$ denote the standard deviation of the
$\widetilde{L}$ and $\widetilde{E}$ distributions, respectively. The
parameter $\alpha=0.125$ was chosen as a compromise so as to obtain a
locally smooth function without significantly changing the overall
shape of the DF.

The distribution in \EL{} space is not uniform and satellites are most
likely to have values around the peak of the DF,
$\EL\approx(-1.5,0.5)$, which corresponds to the dark coloured region
in Fig. \ref{fig:ELDist}. The \EL{} distribution is bounded on the
lower right hand side by circular orbits. Moving perpendicularly away
from this boundary, the orbits become increasingly radial. The
$\widetilde{E}$ distribution is bounded by the potential energy of the
inner radial cut, and the $\widetilde{L}$ distribution is bounded by a
circular orbit at the outer radial cut. In our sample, approximately
$1\%$ of the satellites are unbound, i.e. $E>0$, which is consistent
with previous studies \citep{BK13ApJ...768..140B}. However, we note
that we do not calculate the exact binding energy of each satellite,
but only an approximate value under the assumption that the host halo
is spherically symmetric and well described by an NFW profile (see
eq. \ref{eq:NFWPotEq}). While not shown in Fig.~\ref{fig:ELDist}, we
do keep unbound satellites in our analysis and thus we make no
explicit assumption that MW satellites, such as Leo I, are
bound. Instead, it is simply improbable that Leo I is unbound, and
this is reflected in the individual satellites mass estimates we
present in Section \ref{sec:MW}.

There are several advantages to obtaining a composite DF that is
averaged over many host haloes instead of calculating individual
distributions for each halo, as done by
\citet{Li2017ApJ...850..116L}. In \eagle{}, the mass resolution limits
the number of subhaloes that can be identified in each system. As a
result, the satellite population of each system represents a poor
sampling of their haloes unique DF. The total composite DF contains
many possible halo histories, and their multiplicity effectively
serves as a prior probability. With further knowledge of the MW's
assembly history, it would be possible to restrict the model sample to
have similar assembly histories to the MW. This could reduce the
effective halo scatter and potentially result in a more accurate mass
estimate. However, in this work we choose not be too restrictive.
 
\section{Method}
\label{sec:method}
We present three different methods for inferring the mass of the MW,
each based on the following satellite properties: i)~orbital angular
momentum, ii)~orbital energy, and iii)~both angular momentum and
energy. All three methods employ the same principles and steps. We
focus the discussion on the third method, which combines both $L$ and
$E$, and which should give the best mass constraints since it uses the
largest amount of information. The methods we use are based on the
approach of \citet{Li2017ApJ...850..116L}, which we have modified to
work with a large sample of haloes and \change{ our mass independent DF, \FEL{}.}

We are interested in determining the mass of a host halo starting from
the observed position and velocities of a set of
$N_{\textnormal{Sat}}$ satellites. Each satellite, $s$, has a set of
observed phase-space coordinates:  
\begin{equation}
	\boldsymbol{x}^{s}=\left(v^{s},v_{\textnormal{t}}^{s},r^{s}\right)\\
	\left\{ \boldsymbol{x}^{s}\right\} _{s\in\left[0,N_{\textnormal{Sat}}\right]}
    \;,
\end{equation}
consisting of the speed, $v$, the tangential velocity component,
$v_{\textnormal{t}}$, and the distance, $r$, from the host centre. These
properties, combined with assumptions about the mass, $M_{200}$, and
the 
density profile of the host, are sufficient to calculate the energy
and angular momentum, $\left.\left\{
    \widetilde{E}^{s},\widetilde{L}^{s}\right\} \right|_{M_{200}}$,
of each satellite. Varying $M_{200}$ gives a path in the \EL{}
plane for each satellite.  As a function of $M_{200}$, \Ltilde{}
scales as $M_{200}^{-2/3}$ and so decreases asymptotically to zero
for increasing value of $M_{200}$. The scaled energy, \Etilde{}, has
two terms that scale differently; the kinetic term scales as
$M_{200}^{-2/3}$, while the potential term scales as
$M_{200}^{1/3}$. With increasing $M_{200}$, the potential term
dominates and \Etilde{} tends to $-\infty$.  

\begin{figure}
	\includegraphics[width=\columnwidth]{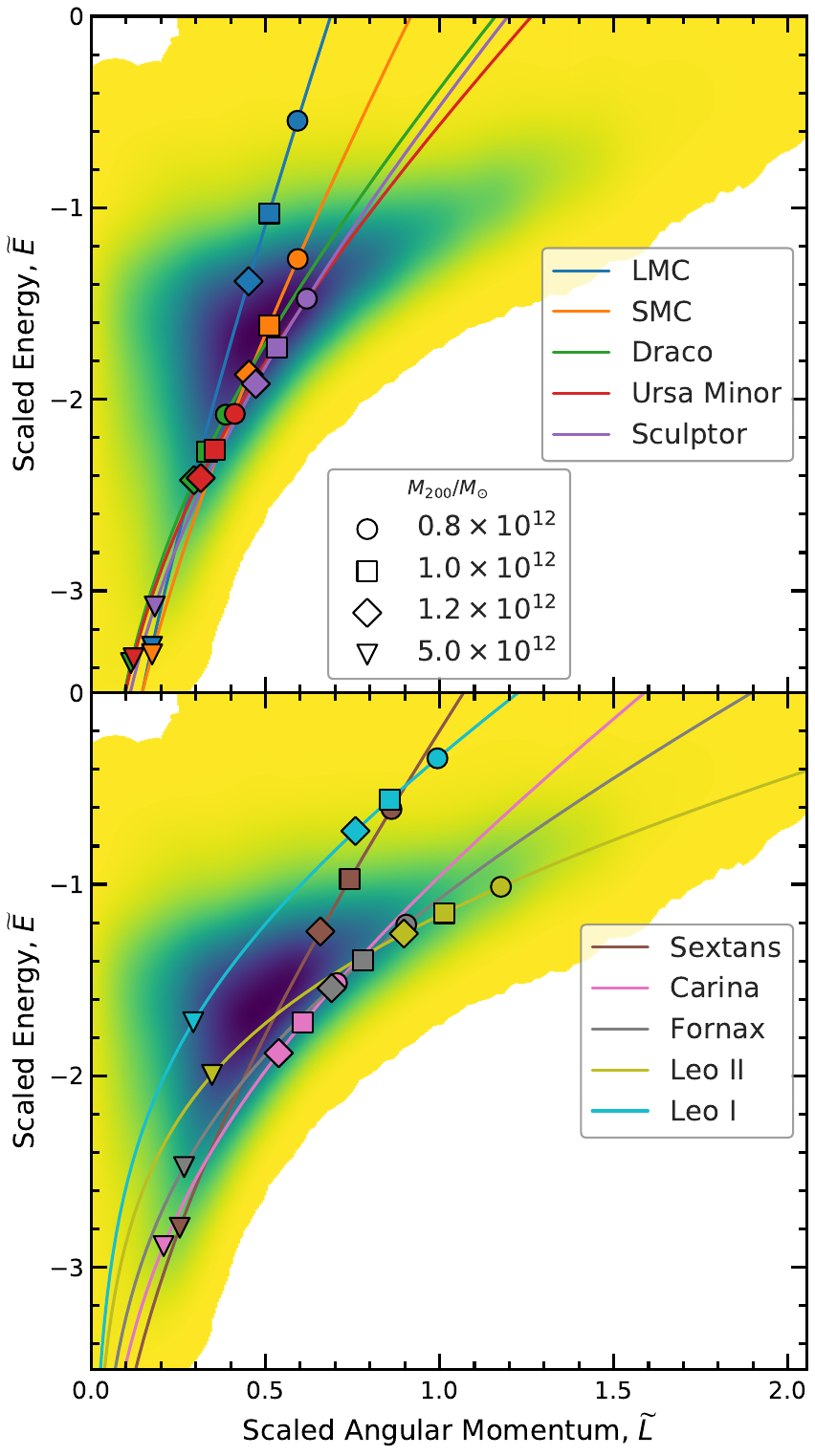}
    \smallspace{}
    \caption{The path of the Galactic satellites in scaled energy --
      angular momentum space, \EL{}, when varying the \MW{} halo mass,
      $M_{200}$. Each curve corresponds to a different satellite (see
      legend). The filled symbols show the location corresponding to
      the four values of $M_{200}$ given in the legend. The energy has
      been calculated using an NFW profile with a concentration of
      8. The colour scheme is the same as in Fig. \ref{fig:ELDist},
      with darker colours corresponding to higher number densities.  }
    \label{fig:MWPaths}
\end{figure}

Fig. \ref{fig:MWPaths} illustrates the path of the Galactic satellites
in the \EL{} plane as we vary the assumed mass of the MW halo. For
example, as we increase the value of $M_{200}$, the LMC dwarf moves
from the top part of the plot to the bottom-left corner. This is
because both \Ltilde{} and \Etilde{} decrease with increasing
$M_{200}$ values.

\change{The trajectory of the satellites through the 2D plane depends on the satellites' orbital phase. The scaled angular momentum, \Ltilde, varies as a function of $M_{200}$ uniformly throughout the orbit, but the rate of change of the scaled energy, \Etilde, is dependent on the satellites' current radius. Nearer pericentre, the satellites have higher absolute values of kinetic and potential energy components compared to those at larger distances. When increasing $M_{200}$, the scaled kinetic energy decreases while the absolute value of the scaled potential energy increases, causing the total scaled energy, \Etilde, to decrease rapidly and thus results in a more vertical trajectory.  The figure also illustrates that when the assumed $M_{200}$ is very high, $\Ltilde{}$ varies slowly and so the paths become nearly vertical.}

Fig. \ref{fig:MWPaths} illustrates how the energy and angular momentum
of satellites can be used to determine the host halo mass. The DF in
\EL{} space is not uniform, and as the assumed $M_{200}$ of the host
is varied, satellites move between regions of high and low number
density in this space. For example, the LMC falls in a high density
region for $M_{200}\approx1.4\times10^{12}\Msun$, and in lower density
regions for higher or lower masses. Thus, the LMC phase space
coordinates would prefer a MW halo mass of
$\approx1.4\times10^{12}\Msun$. In contrast, the Leo I path is nearest
to the maximum density for $M_{200}\approx2.9\times10^{12}\Msun$, and
suggests a higher MW mass.

We now describe how each satellite can be used to obtain a likelihood
for the MW halo mass, and how to combine the mass estimates from
various satellites. Our aim is to determine the likelihood,
$p(M_{200}|\mathbf{x}^s)$, for the host mass given the observed
$\mathbf{x}^s$ properties of satellite $s$. 

\change{The likelihood can be calculated from the \Etilde{} distribution via
\begin{equation}
	p(M_{200}|\mathbf{x}^s) = \FEE{} ~ \frac{\partial\Etilde}{\partial M_{200}} \bigg|_{\Etilde=\Etilde^s}
   	\label{eq:mass_likelihood1d}
    \;,
\end{equation}
where the \FEE{} term denotes the DF, while the partial derivative arises from the Jacobian of the transformation from \Etilde{} to host halo mass, $M_{200}$. 
The same procedure can be used to estimate the host mass using only the angular momentum by replacing \FEE{} by the \Ltilde{} distribution function, \FLL{}, and by changing the \Etilde{} derivative term to \Ltilde{}, to obtain:
\begin{equation}
	p(M_{200}|\mathbf{x}^s) = \FLL{} ~ \frac{\partial\Ltilde}{\partial M_{200}} \bigg|_{\Ltilde=\Ltilde^s}
   	\label{eq:mass_likelihood1d_L}
    \;.
\end{equation}
This expression can be extended to the 2 dimensional case, where we use both \EL{} to constrain the halo mass, via
\begin{equation}
	\begin{split}
	p(M_{200}|\mathbf{x}^s) = F(\Etilde,\Ltilde) ~ M_{200} ~ \frac{\partial\Etilde}{\partial M_{200}} \frac{\partial\Ltilde}{\partial M_{200}}\bigg|_{\Etilde=\Etilde^s;\; \Ltilde=\Ltilde^s}
   	\label{eq:mass_likelihood2d}
    \;,
    \end{split}
\end{equation}
where the additional $M_{200}$ factor is needed to have the correct units. Note that all the \Etilde{} and \Ltilde{} terms in Eqs. \eqref{eq:mass_likelihood1d}-\eqref{eq:mass_likelihood2d} are evaluated at the point $\Etilde^{s} \equiv\Etilde(\mathbf{x}^s,M_{200})$ and  $\Ltilde^{s} \equiv\Ltilde (\mathbf{x}^s,M_{200})$.  For a detailed derivation of Eqs. \eqref{eq:mass_likelihood1d}-\eqref{eq:mass_likelihood2d} please see Appendix \ref{appendix:stats}.
}

In practice, we actually determine the logarithm of the mass, $\log_{10}(M_{200})$, since the resulting probability distribution function (PDF) in log space is closer to a Gaussian. 
We determine the most likely host mass as the mass that maximizes the likelihood --- the Maximum Likelihood Estimator (MLE) mass, $M_{200}^{\textnormal{MLE}}$. As the uncertainties, we take the 68\% confidence limits corresponding to the interval between the 16 and 84 percentiles of the mass PDF. 
Assuming that the satellites are independent tracers, we can combine
the estimates for individual satellites to obtain an overall estimate
given a set of observations, $\{ \boldsymbol{x}^{s}\}$. The combined
likelihood is given by: 
\begin{equation}
	p\left(M_{200}\left|\left\{ \boldsymbol{x}^{s}\right\} \right.\right) = 
	\underset{s=1}{\overset{N_{\textnormal{Sat}}}{\prod}}p\left(M_{200}\left|\boldsymbol{x}^{s}\right.\right)
    \;.
\end{equation}

The potential energy of satellites has a weak dependence on the host
halo concentration, which is an unknown quantity. We have tested that
the 10 satellites used here cannot, by themselves, place any meaningful
constraints on the concentration of the MW halo. Thus, we proceed to
marginalize over the unknown concentration:
\begin{equation}
	p\left(M_{200} \left|{\boldsymbol{x^{s}}} \right.\right) = \intop p\left(M_{200}\left|{\boldsymbol{x^{s}}},C\right.\right)p\left(C\left|M_{200}\right.\right)\textnormal{d}C 
    \;,
\end{equation}
where $p(C|M_{200})$ denotes the distribution of concentrations
for haloes of mass, $M_{200}$, found in the \eagle{} simulation, which
we took from \citet{MatthieuNFWConc2015MNRAS.451.1247S}. In practice,
we evaluate $ p(M_{200}|\boldsymbol{x},C)$ using 15  evenly
spaced values in the range $C\in\left[5,20\right]$. We note that the
dependence on concentration is weak, so our results are not affected
by the choice of the distribution of concentrations (see Appendix~\ref{Appendix:Concentration})  
 
\subsection{Observational errors}
\label{sec:method_obs_errors}
While we have perfect knowledge of the phase space coordinates,
$\{\boldsymbol{x}^{s}\}$, of \eagle{} satellites, in order to apply
the method to the MW satellites we must consider the effects of
observational errors. To account for errors, we perform a set of 1000
Monte Carlo realizations that sample the observational uncertainties
(see Section~\ref{Sec:ObsData} for a detailed description of the
procedure). This produces a Monte Carlo sample of allowed phase-space
coordinates for each satellite. We first determine the MW mass
likelihood for each Monte Carlo realization, and then we average the
likelihood of all the Monte Carlo samples. In the limit of a large
number of Monte Carlo samples, this is equivalent to marginalizing
over the observational errors.

\subsection{Method calibration using EAGLE}
\label{SubSec:EAGLETesting}

To provide a robust mass estimate of the MW halo, we now explore the
accuracy of our methods using tests on mock satellite systems. Since
MLE estimates can be biased, we first calibrate the inference methods
using a large sample of \eagle{} systems. Then, in Section
\ref{Sec:auriga_tests}, we validate the methods on an independent,
higher resolution set of simulations taken from the \auriga{} project.

\begin{figure} 
	\includegraphics[width=\columnwidth]{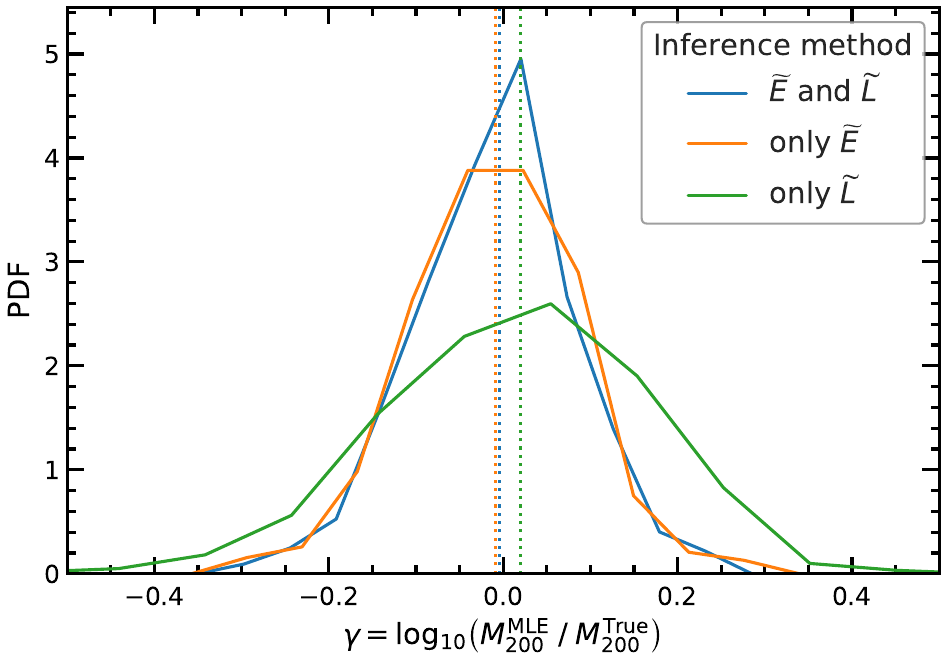} 
    \smallspace{} 
    \caption{The distributions of the ratio of MLE estimate,
      $M_{200}^{\textnormal{MLE}}$, to the true halo mass
      $M_{200}^{\textnormal{True}}$, from each of our three
      inference methods. The results were obtained by applying 
      each mass determination method to a sample of
      $\sim600$ systems from the \eagle{} simulation. The vertical
      dotted lines indicate the median of each distribution, which
      represents the bias, $b$, of each method. For subsequent
      results, we correct the mass estimates by the bias of each
      method and we denote the corresponding mass by
      $M_{200}^{\textnormal{Esti}}$.} 
	\label{fig:EagleTestingRatio} 
\end{figure}

To calibrate the three mass determination methods we start by
applying them to the \eagle{} simulations. We select the same \eagle{}
haloes as in Section \ref{Sec:EagleSims}, that is haloes of 
total mass ${\sim}10^{12}\Msun$, and keep only those which contain at
least 10 luminous satellites within the distance range quoted in
Section~\ref{Sec:EagleSims}. There are ${\sim}600$ haloes satisfying
the selection criteria. We then apply each mass determination method
to each \eagle{} system to obtain the MLE mass,
$M_{200}^{\textnormal{MLE}}$ of that system. The results are shown in
Fig. \ref{fig:EagleTestingRatio}, where we compare the MLE masses to 
the true total halo mass, $M_{200}^{\textnormal{True}}$.  The performance
of each method may be quantified by the ratio,
$\gamma=\log_{10}\left(M_{200}^{\textnormal{MLE}}/M_{200}^{\textnormal{True}}\right)$,
for each \eagle{} system. The median and scatter of the $\gamma$
distribution give the bias and typical uncertainty of the method,
respectively.

\change{Fig. \ref{fig:EagleTestingRatio}  shows that our three methods have only small biases compared to their dispersion. The $(E,L)$ and the E only methods have a slight bias with the median
of the $\gamma$ distribution being $-0.01$, while the method based on $L$ only has an bias of $+0.02$.} A consistently biased estimate is not a problem since it can easily
be corrected to obtain an accurate result. The bias-corrected mass
estimate, $M_{200}^{\textnormal{Esti}}$, is given by: 
\begin{equation}
	\log_{10}\left(M_{200}^{\textnormal{Esti}}\right)=\log_{10}\left(M_{200}^{\textnormal{MLE}}\right)-b
    \label{eq:bias_correction} \;.
\end{equation}

The dispersion of the $\gamma$ distributions in
Fig.~\ref{fig:EagleTestingRatio} reflects the true precision of the
method, $\sigma_{\textnormal{True}}$. Mass estimates based only on the
angular momentum have the largest dispersion, 
$\sigma_{\textnormal{True}}=0.15$, while both $E$ and $(E,L)$ methods
have the same precision, $\sigma_{\textnormal{True}}=0.09$.	Thus,
most of the mass information is contained in the satellites' orbital
energy. Adding angular momentum data hardly improves the mass
estimates, indicating that $L$ does not contain significant
information about the host mass beyond the information already
contained in the satellites' energy. 

Another important point to consider is the confidence interval to be 
associated with each mass measurement. One possibility is to take the
dispersion of $\gamma$ (see Fig.~\ref{fig:EagleTestingRatio}), but
this suffers from the limitation of assigning the same error to all mass
measurements. In practice, the mass of some host haloes can be more
precisely determined than the mass of others, and the confidence limits do
not need to be symmetrical around the MLE value (e.g. see
Fig.~\ref{fig:AurigaTestingRatio}). Thus, the approach of assigning
a single error to all measurements is not optimal. 

An alternative is to consider the error estimates of the Bayesian
method. These should be accurate, except for the effects of an
assumption implicit in our method, that all satellites are independent
tracers. For example, satellites can fall in groups or filaments,
which might result in correlated energy and angular momentum amongst
two or more satellite galaxies. For the brightest 10 satellites, the
ones considered here, only a small fraction is expected to have fallen
in groups \citep[e.g.][]{Wetzel2015, Shao2018a} and, in any case,
interactions with other satellites and with the host halo and galaxy
are expected to decrease any phase-space correlations present at the
time of accretion \citep[e.g.][]{Deason2015, Shao2018c}. Thus, we
would generally expect the assumption of independent tracers to be
reasonable. We have checked how realistic the Bayesian error estimates
are and found them to be roughly the same as the uncertainties shown in Fig.~\ref{fig:EagleTestingRatio}. The same will not hold true in
future studies when the method will be applied to much larger numbers of satellites
(see discussion in Section \ref{SubSec:Improving}).

\subsection{Tests with the AURIGA simulations}
\label{Sec:auriga_tests}

In this section we test our mass inference methods by applying them to
model galaxies from the \auriga{} project. \auriga{} is a suite of
high-resolution, hydrodynamical zoom-in simulations of MW-like
systems. We consider the $30$ level 4 systems, which have dark matter
and gas mass resolution ${\sim}30$ times higher than \eagle{} (see
\citealt{Auriga_2017MNRAS.467..179G} for details).  \auriga{} makes
for a perfect test suite since it has higher resolution, uses a
different hydrodynamics code and includes a different galaxy formation model than \eagle{}. Thus, by applying our inference methods to these
completely independent simulations, we can assess our
methods' accuracy and quantify any systematic biases that may have
been introduced by calibrating our methods on the \eagle{}
simulations.

\begin{figure}
	\includegraphics[width=\columnwidth]    {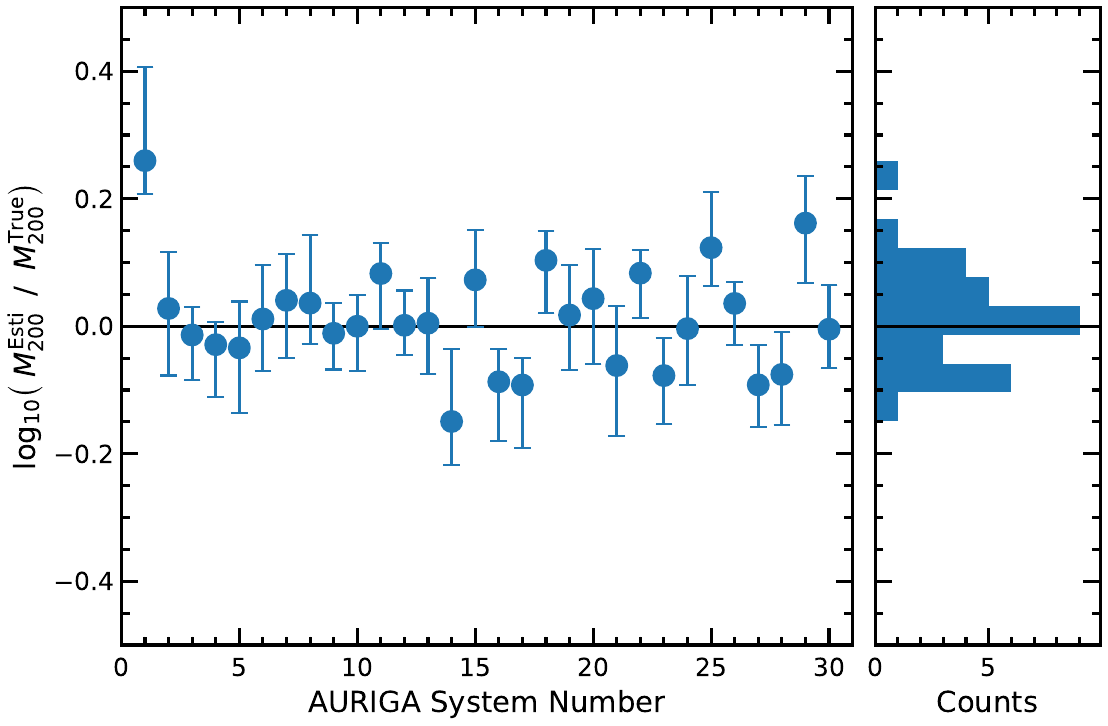}
    \smallspace{}
    \caption{Test of the energy -- angular momentum halo mass
      inference method on 30 MW-mass galaxies from the \auriga{}
      galaxy formation simulation project. We show the ratio between
      the estimated, $M_{200}^{\textnormal{Esti}}$, and the true,
      $M_{200}^{\textnormal{True}}$, halo masses for each \auriga{}
      system. Note that $M_{200}^{\textnormal{Esti}}$ includes the bias correction determined from the \eagle{} mock catalogues (see equation \ref{eq:bias_correction}). 
      The errorbars correspond to the estimated $68\%$
      confidence limit. The \auriga{} simulations have much higher
      resolution and assume different galaxy formation models than
      \eagle{}, and thus provide a rigorous test of the mass inference
      method. Most mass estimates agree with the true values within
      the $68\%$ confidence limit, in very good agreement with
      statistical expectations.  }
    \label{fig:AurigaTestingRatio}
\end{figure}

For each \auriga{} galaxy, we identify the brightest 10 satellites
galaxies at a distance between $40$ and $300\kpc$ from the halo
centre. These objects represent our mock observational sample of the
MW-like satellite systems. We then apply the $(E,L)$ mass
determination method to each of the 30 \auriga{}
systems. 

Fig. \ref{fig:AurigaTestingRatio} shows the ratio of estimated to true
masses, as well as the associated uncertainties for each \auriga{}
galaxy. We find that for 19 out of the 30 systems, or $63\%$, the
estimated mass agrees with the true value to the 68\% confidence
interval, approximately as expected from the statistics. \change{This performance is very good especially when taking into account that around a third of the \auriga{} systems are unrelaxed (see Sec. \ref{Sec:EagleSims} for relaxation criteria).}
We have checked
that the other two methods, using only $L$ and only $E$, are similarly
successful. This test demonstrates the accuracy of our method for
determining halo masses and confirms that our error estimates are
realistic and robust.  

\section{ Milky Way Mass Estimates}
\label{sec:MW}

\begin{figure}
	\includegraphics[width=\columnwidth]{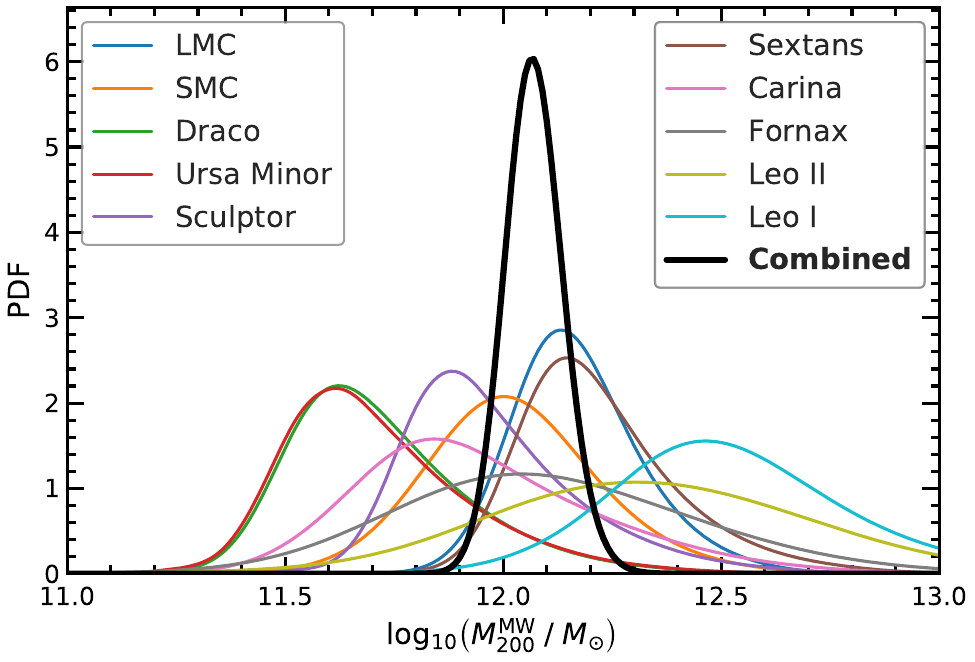}
    \smallspace{}
    \caption{ The MW halo mass, $M_{200}^{\mathrm{MW}}$, inferred from
      the energy and orbital angular momentum of each classical
      satellite (except Sagittarius). The thick line shows the
      inferred MW halo mass,
      $M_{200}^{\mathrm{MW}}=1.04_{-0.14}^{+0.23}
      \times10^{12}M_\odot$ (68\% confidence limit), obtained by
      combining the 10 individual estimates. The inferred
      $M_{200}^{\mathrm{MW}}$ values and their corresponding errors
      are given in Table~\ref{Tab:MassResults}.  }
    \label{fig:MWSatsPdfsSinglePlot}
\end{figure}

We now apply our mass estimation methods to data for the 10 MW
satellites that satisfy our selection criteria. We begin by obtaining
the Galactic halo mass likelihood from each satellite and
corresponding uncertainties (calculated with the Monte Carlo sampling
technique described in Section~\ref{sec:method_obs_errors}). The PDFs
of the MW halo mass, $M_{200}$, obtained from each satellite's data
using the $(E,L)$ method are shown in
Fig.~\ref{fig:MWSatsPdfsSinglePlot}; the best estimates and associated
68\% confidence intervals are given in Table~\ref{Tab:MassResults}.

\begin{table}
	\centering
	\caption{ The MW halo mass, $M_{200}^{\mathrm{MW}}$ (the mass enclosed within a sphere of average density $200$ times the critical density) estimated from each classical satellite (except Sagittarius), and the combined overall result. The table gives mass estimates using: (i)~only the angular momentum, $L$; (ii)~only the energy, $E$: and (iii)~both $E$ and $L$. We quote $68\%$ confidence limits.
    }
    \label{Tab:MassResults}
    \setlength\extrarowheight{5pt}
\begin{tabularx}{0.74\columnwidth}{l r r r}
\hline
& \multicolumn{3}{c}{ $M_{200}^{\mathrm{MW}} \; [10^{12}M_{\odot}]$
}\tabularnewline
Satellite & only $L$ & only $E$  &  $E$ and $L$ \tabularnewline
\hline
LMC & $0.98_{-0.51}^{+1.78}$ & $1.23_{-0.25}^{+0.65}$ & $1.35_{-0.28}^{+0.76}$\tabularnewline
SMC & $0.98_{-0.52}^{+1.84}$ & $0.93_{-0.31}^{+0.61}$ & $1.00_{-0.32}^{+0.68}$\tabularnewline
Draco & $0.51_{-0.26}^{+0.94}$ & $0.4`_{-0.09}^{+0.39}$ & $0.42_{-0.08}^{+0.43}$\tabularnewline
Ursa Minor & $0.56_{-0.29}^{+1.03}$ & $0.40_{-0.09}^{+0.40}$ & $0.42_{-0.09}^{+0.43}$\tabularnewline
Sculptor & $1.02_{-0.52}^{+1.88}$ & $0.74_{-0.15}^{+0.66}$ & $0.76_{-0.14}^{+0.74}$\tabularnewline
Sextans & $1.70_{-0.87}^{+3.09}$ & $1.35_{-0.29}^{+1.01}$ & $1.41_{-0.28}^{+1.12}$\tabularnewline
Carina & $1.29_{-0.69}^{+2.34}$ & $0.74_{-0.24}^{+0.83}$ & $0.69_{-0.21}^{+1.02}$\tabularnewline
Fornax & $1.86_{-1.08}^{+3.63}$ & $1.12_{-0.52}^{+1.68}$ & $1.10_{-0.52}^{+1.78}$\tabularnewline
Leo II & $3.02_{-1.86}^{+5.63}$ & $1.91_{-1.01}^{+4.32}$ & $2.04_{-1.11}^{+3.17}$\tabularnewline
Leo I & $2.40_{-1.49}^{+4.61}$ & $3.09_{-1.16}^{+6.45}$ & $2.88_{-1.06}^{+3.43}$\tabularnewline[0.1cm]
\hline 
\textbf{Combined} & $\mathbf{ 1.20_{-0.27}^{+0.42} }$ & $\mathbf{ 1.10_{-0.14}^{+0.21} }$ & $\mathbf{ 1.17_{-0.15}^{+0.21} }$\tabularnewline[0.1cm]
\hline
\end{tabularx}
\end{table}

Individually, the satellites give a wide range of total masses for the
MW. For example, Ursa Minor and Draco favour a very low mass,
$M_{200}\approx 10^{11.6}\Msun$, which is because both of them have very low total specific energies (see Table \ref{Tab:SatObs}).
At the other extreme, Leo~I has the highest total energy and  
favours a halo an order of magnitude more massive,
$M_{200}\approx 10^{12.5}\Msun$. However, the mass estimate from any
one satellite has a broad distribution and does not provide a strong
constraint on the MW mass. The true power of the method comes from
combining the mass likelihoods from each satellite; the combined
result is shown as a thick line in Fig~\ref{fig:MWSatsPdfsSinglePlot}. The
combined estimate for the MW halo mass is 
$M_{200}^{\mathrm{MW}}=1.17_{-0.15}^{+0.21} \times10^{12}M_\odot$.

\begin{figure}
	\includegraphics[width=\columnwidth]{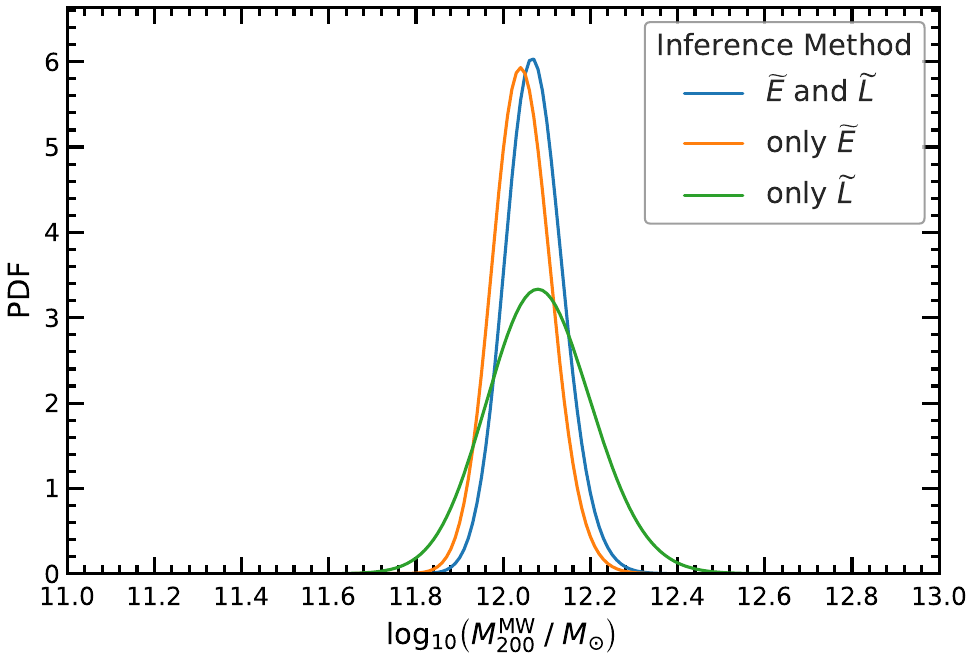}
    \smallspace{}
    \caption{ Comparison of the MW halo mass inferred using the three
      methods studied here. The methods use the following satellite
      data: (i)~only the angular momentum, $L$; (ii)~only the energy,
      $E$; and (iii)~both $E$ and $L$. The mass estimates and their
      errors are given in Table~\ref{Tab:MassResults}. }
    \label{fig:MWCombinedPdf}
\end{figure}

\begin{figure*}
	\includegraphics[width=0.65\textwidth]{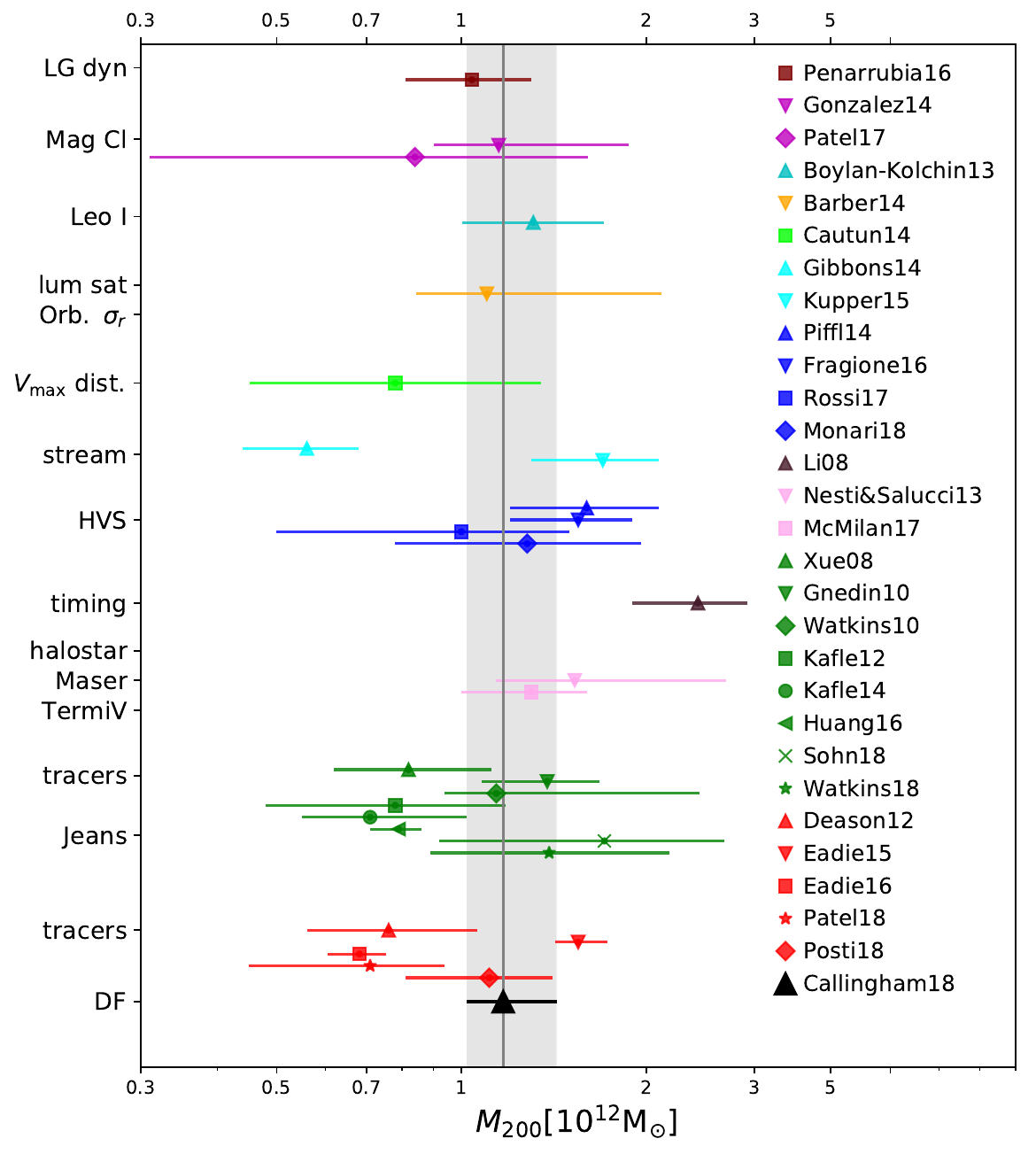}
    \vspace{-.3cm}
    \caption{ Comparison of our inferred MW halo mass with a selection
      of previous estimates. The vertical line and the shaded region
      show our $M_{200}$ estimate and its $68\%$ confidence limit. The
      remaining symbols show previous estimates (see legend), with
      the horizontal lines corresponding to the quoted $68\%$
      confidence limits. The results are grouped according to the
      methodology employed (see vertical axis). We give the mass,
      $M_{200}$, contained within $R_{200}$ (the radius
      enclosing a mean density equal to 200 times the critical
      density). Some of the previous estimates were converted to
      $M_{200}$ by assuming an NFW profile and the mean concentration
      predicted for that mass. 
	}
    \label{fig:LitComparison}
\end{figure*} 

Fig.~\ref{fig:MWCombinedPdf} compares the Galactic halo mass
determination using the three methods introduced in this study. We
find very good agreement amongst the three, with all of them having
a very large overlap (see Table~\ref{Tab:MassResults} for the actual
values and their uncertainties). Of the three, the method based on
angular momentum only is the most uncertain and, of the remaining
two, the one based on energy only gives a slightly lower
uncertainty. As we saw in Fig.~\ref{fig:EagleTestingRatio}, adding $L$
data to $E$ data does not produce an improvement in the mass
determination, which is what we find here too. In fact, the $(E,L)$
method seems to have slightly larger uncertainties than the $E$-only
method; however, the difference is very small and not statistically
significant. We also find that the estimated uncertainties in the MW
mass determination are similar to the ones shown in
Fig.~\ref{fig:EagleTestingRatio}, where we tested the methods on the
\eagle{} simulations. As we will see in Figures~\ref{fig:MWObs_Vary}
and~\ref{fig:NSat_Tracers}, the uncertainties in the mass are dominated by
the small number of satellites, not by their proper motion errors.

It is important to consider possible systematics that may affect our
mass determination. For example, the LMC and SMC are believed to have
fallen in recently as a pair \citep[e.g.][]{Kallivayalil2013}, and
might not encode independent information about the MW halo. We have
checked that discarding the SMC from our analysis does not
significantly change the median estimate and leads only to a small
increase in the uncertainty range. We also know that the classical
satellites are atypical in at least two respects: they currently
reside in a thin plane, with several orbiting preferentially within
it, and they have a very low velocity anisotropy. These two properties
place the MW satellite system in the tail of the $\Lambda$CDM
expectations \citep[e.g. see][]{Pawlowski2014,
  Cautun2015,Cautun2017}. The analysis described in
Appendix~\ref{Appendix:maximum_likelihood} shows that the distribution
of $E$ and $L$ values of the Galactic satellites is, in fact,
consistent with $\Lambda$CDM predictions, with no evidence for any
tension.

\subsection{Comparison to previous MW mass estimates}
\label{sec:previous_work}

In Fig.~\ref{fig:LitComparison} we compare our total MW halo mass
estimate with a selection of results from previous studies. This
figure is an update of Figure~1 in
\cite{Wang15_DMHaloDynTracers_MNRAS.453..377W} and includes recent
estimates, especially those that use \textit{Gaia}~DR2 data. Some mass
determination methods, such as ours and those based on Local Group
dynamics \citep[e.g.][]{Li2008,Penarrubia2016} and satellite
dynamics \citep[e.g.][]{Watkins2010,BK13ApJ...768..140B,Barber2014,Eadie2015},
give the total mass directly, but many others, such as those using globular
clusters \citep[e.g.][]{PostiHelmi2018arXiv180501408P,
  WatkinsMWGC2018arXiv180411348W} or halo stars \cite[e.g.][]{Xue2008,Gnedin2010,
  Deason12_BrokenDegenMNRAS.424L..44D,Huang2016}, give the enclosed mass only
within an inner region of the MW halo and require an assumption about
the MW halo mass profile for extrapolation to the total mass.  Despite
the wide range of values quoted in the literature, our result is
consistent within $1\sigma$ with the majority of previous mass
estimates. Our errors are significantly smaller than those of most
previous estimates and, most importantly, we have rigorously and
extensively tested our method on simulated galaxies to produce an
accurate, unbiased mass estimate with realistic uncertainties.

Our estimated value of ${\sim} 10^{12}\Msun$ for the MW halo mass has
important implications for the interpretation of the satellite
population of our galaxy, which is often used as a testbed for the
$\Lambda$CDM model. For example, the ``too-big-to-fail" problem
\citep{BK11_TBtF_MNRAS.415L..40B}, which refers to the number of
massive, dense satellites in the MW halo, is significantly
alleviated. Indeed, \cite{Wang12_MissingSat_MNRAS.424.2715W}
showed that approximately 40\% of haloes with mass
$M_{\rm halo} \sim 10^{12}\Msun$ in $\Lambda$CDM dark matter only
simulations have three or fewer subhaloes with $V_{\rm max} > 30$ km/s
(the threshold used by \citealt{BK11_TBtF_MNRAS.415L..40B} to define
massive failures). For the MW halo mass that we infer, the
``too-big-to-fail problem'' is not a failure of $\Lambda$CDM.

An accurate estimate of the MW halo mass is also crucial in order to
address properly the missing satellites problem. The total number of
subhaloes depends strongly on the halo mass (doubling the halo mass,
roughly doubles the number of subhaloes). Thus, when appealling to
baryonic physics solutions to this problem, such as the influence of
reionization and stellar feedback, an accurate estimate of the halo
mass is a pre-requisite for a realistic model. Moreover, when the halo
mass is known, the number of subhaloes may even inform us about these
critical processes, such as when the epoch of reionization occured (see
e.g. Figure~1 in \citealt{Bose2018}), or indeed about the identity of
the dark matter
\citep{Kennedy2014:WDMParticlefromMWSat_MNRAS.442.2487K,Lovell2014:WarmDM_MNRAS.439..300L}. 

\subsection{The concentration of the MW halo}
\label{sec:concentration}

\begin{figure}
	\includegraphics[width=\columnwidth]{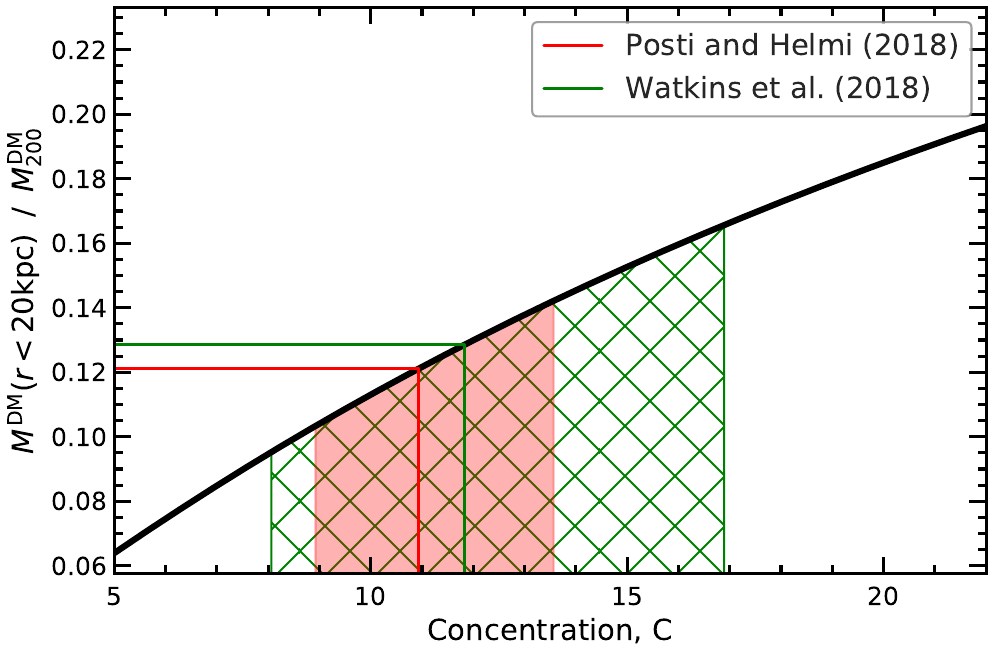}
    \smallspace{}
    \caption{The concentration of the MW halo inferred by combining
      our total mass estimate with previous inner mass estimates. The
      solid thick curve shows the dark matter (DM) mass fraction,
      $M^\textnormal{DM}(<20~\rm{kpc})/M^\textnormal{DM}_{200}$,
      contained within $20~\rm{kpc}$ of the halo centre as a function
      of concentration, $C$, for our best estimate of a total halo
      mass of $M^\textnormal{DM}_{200}=1.11\times10^{12}M_\odot$.  The
      two horizontal lines correspond to the
      \citet{PostiHelmi2018arXiv180501408P} and
      \citet{WatkinsMWGC2018arXiv180411348W} inner mass estimates. The
      inferred concentrations are shown by the two vertical lines,
      with the shaded regions corresponding to the $68\%$ confidence
      ranges. We find $C=10.9^{+2.6}_{-2.0}$ and
      $C=11.8^{+5.1}_{-3.8}$ for the
      \citeauthor{PostiHelmi2018arXiv180501408P} and
      \citeauthor{WatkinsMWGC2018arXiv180411348W} inner mass
      estimates, respectively.}
    \label{fig:ConcInfer}
\end{figure}

\begin{figure}
	\includegraphics[width=\columnwidth]{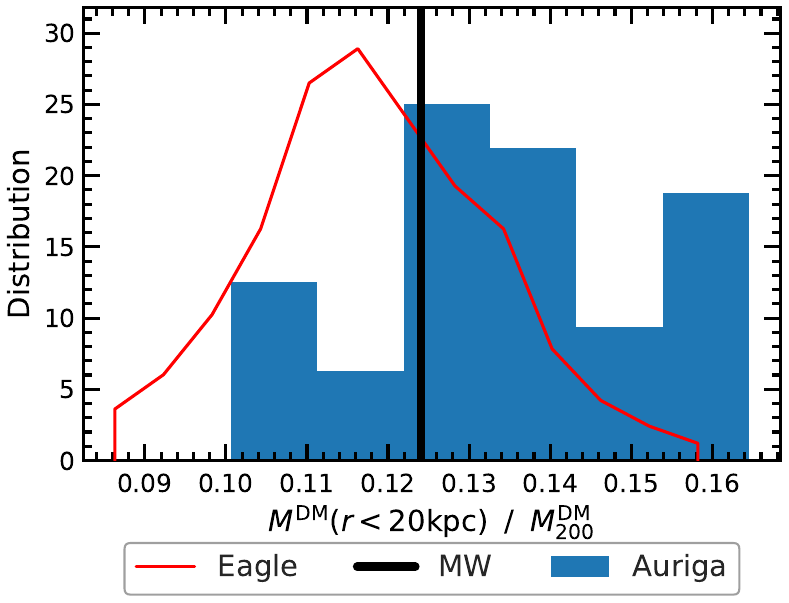}
    \smallspace{}
    \caption{\change{The distribution of the DM mass fraction contained within $20~\rm{kpc}$ of the halo centre, $M^\textnormal{DM}(<20~\rm{kpc})/M^\textnormal{DM}_{200}$, for MW-sized galaxies in \eagle{} and \auriga{}. 
    The red line shows the distribution of systems from our \eagle{} sample described in Section 2.3, whose ${M}_{200}$ is within 0.2 dex of our MW mass estimate. The blue histogram gives the distribution of the 30 level 4 \auriga\ systems described in Section 3.3. The thick black line shows the MW's DM mass fraction; calculated  using our own MW halo mass estimate, $M^\textnormal{DM}_{200}$, and  $M^\textnormal{DM}(<20~\rm{kpc})$ from \citet{PostiHelmi2018arXiv180501408P}.}}
    \label{fig:MasFraction}
\end{figure}

Alongside mass, the other fundamental property of DM haloes is their
concentration. Besides being one of the key parameters of the NFW
profile, the concentration encodes crucial information about the
halo's formation history
\citep[e.g.][]{Wechsler2002,Lu2006,Ludlow2014} and, after halo mass,
is the most important property for determining how galaxies populate
haloes \citep[e.g.][]{Matthee2017}. Our MW halo mass estimate does not depend on, nor constrain, the MW halo concentration. However, 
when combined with mass estimates for the inner regions of the Galaxy, we can use our mass estimate to infer the concentration of the MW halo. For this, we
use inner mass determinations based on the dynamics of the globular
cluster population. This population is much more radially concentrated
than the satellite galaxy population, and there is a large number of
globular clusters with precise \textit{Gaia} DR2 proper motion
measurements \citep{GaiaDR22018arXiv180409365G}. This enabled
\citet{PostiHelmi2018arXiv180501408P} and
\citet{WatkinsMWGC2018arXiv180411348W} to estimate the total mass
enclosed within ${\sim}20\kpc$ from the Galactic Centre with high
precision.

To determine the concentration we assume that the DM distribution
follows the NFW profile, which provides a very good fit to the DM
density profiles in both DM-only and hydrodynamic simulations. To
determine the enclosed DM mass, we subtract the MW baryonic mass,
$M^\textnormal{baryons}_\textnormal{MW}$, from the total mass
measurements within both $20\kpc$ and $R_{200}$. We use the
\citet{McMillan2017} estimates: a stellar mass of
$5.4\times10^{10}\Msun$ and a gas mass of $1.2\times10^{10}\Msun$,
which corresponds to
$M^\textnormal{baryons}_\textnormal{MW} = 6.6\times10^{10}\Msun$.

Fig. \ref{fig:ConcInfer} shows the fraction of DM mass enclosed within
$20\kpc$ of the centre as a function of the halo concentration; the
solid lines and shaded regions indicate the inferred concentrations
and their 68\% confidence ranges. The
\citet{PostiHelmi2018arXiv180501408P} estimate gives a mass ratio,
$M^\textnormal{DM}(<20\kpc)/M^\textnormal{DM}_{200} \approx 0.12$,
which corresponds to a concentration of $C=10.9^{+2.6}_{-2.0}$ (68\%
confidence limits), where the errors were calculated by Monte Carlo
sampling of the uncertainties associated with the inner and total mass
estimates. The same analysis for the
\citet{WatkinsMWGC2018arXiv180411348W} inner mass estimate gives
$M^\textnormal{DM}(<21.1\kpc)/M^\textnormal{DM}_{200} \approx 0.20$,
and a concentration, $C=11.8^{+5.1}_{-3.8}$. To include the
\citeauthor{WatkinsMWGC2018arXiv180411348W} result in
Fig.~\ref{fig:ConcInfer}, we rescaled their mass estimate to a
fiducial distance of $20\kpc$.

We find that the MW halo has a high concentration for its mass, with a
most likely value of $C \sim 10.9$, which could suggest that the MW halo assembled early. The high MW halo concentration is
supported by other studies; for example, the best fit Galaxy model of
\citet{McMillan2017} gives $C=16\pm3$. In the \eagle{} simulations, the
median concentration of a ${\sim}10^{12}\Msun$ halo is ${\sim}8.2$ and only ${\sim}23\%$ of haloes have a concentration higher than $10.9$ which suggests that the MW halo is an outlier. 

\change{However, the presence of  central baryonic components causes a contraction of the very inner region of ${\sim}10^{12}\Msun$ mass haloes, increasing the total mass in the inner region. As a result, the inner region is not well described by an NFW profile, and the inferred concentration is biased high \citep[e.g.][]{MatthieuNFWConc2015MNRAS.451.1247S}. To overcome this limitation, in Fig. \ref{fig:MasFraction} we compare the inner DM mass fraction of the MW to that of similar mass haloes in the \eagle{} and \auriga{} simulations and find that the MW is typical of haloes in both simulations. The systematic difference between the \eagle{} and \auriga{} distribution reflects the stellar mass content of those objects: compared to abundance matching results, galactic mass haloes in \eagle{} have stellar masses that are too low, while equal mass haloes in \auriga{} have stellar masses that are too high. 
}

\subsection{Improving the mass estimate}
\label{SubSec:Improving}
In this section we discuss the limitations of our method and ways of
improving the MW mass estimate. There are two main sources of
uncertainty: statistical, from the finite number of satellites; and
systematic, from halo-to-halo variation. The former can be reduced by
increasing the number of dynamical tracers and/or reducing
observational errors, but the latter cannot be reduced.

We begin by investigating the effect of observational errors on the MW
halo mass determination. The main source of observational
uncertainties are the proper motion measurements. As such, we consider
the effect of varying the errors, $\sigma_{\mu_{\alpha}}^{s}$ and
$\sigma_{\mu_{\delta}}^{s}$, associated with the two components of the
proper motion. For the MW observations these errors vary from
satellite to satellite, from $0.005~\textnormal{mas/year}$ for
Sculptor to $0.039~\textnormal{mas/year}$ for Leo II, with a median of
${\sim}0.018~\textnormal{mas/year}$. For simplicity, here we assume
the same error for all satellites, that is
$\sigma_{\mu_{\alpha}}^{s}=\sigma_{\mu_{\delta}}^{s}=\sigma_{\mu}$, and study the effect of observational errors by varying $\sigma_{\mu}$. For each $\sigma_{\mu}$ value, we
proceed by taking the current proper motions of each MW satellite and
resetting their errors to the target value of $\sigma_{\mu}$. Then, we
generate a sample of Monte Carlo realizations using the procedure
described in Section~\ref{Sec:ObsData} and apply the mass estimation
method.

\begin{figure}
	\includegraphics[width=\columnwidth]{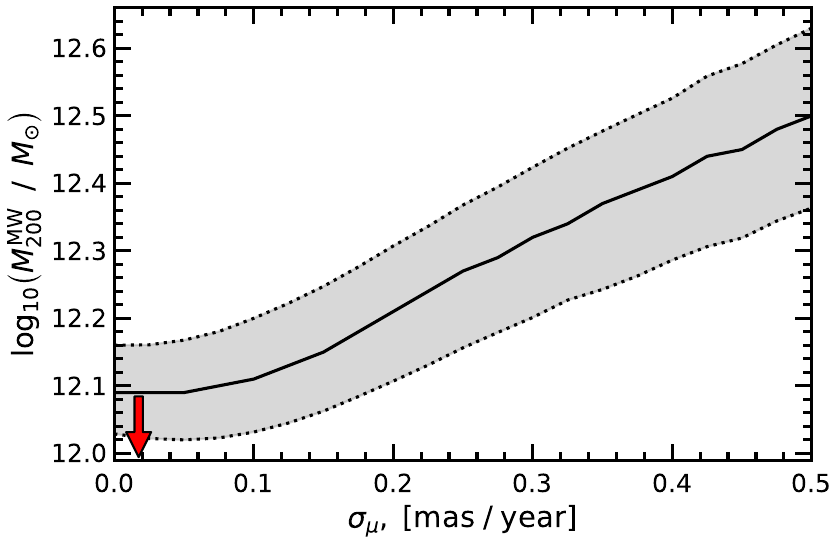}
    \smallspace{}
    \caption{The estimated MW halo mass, $M_{200}^{\textnormal{MW}}$,
      as a function of the size of proper motion errors,
      $\sigma_\mu$. Results are shown only for the inference method 
      based on both $\widetilde{E}$ and $\widetilde{L}$
      values. The solid line gives the mass estimate while the shaded
      region shows the $68\%$ confidence interval. Larger values of 
      $\sigma_\mu$ result in more uncertain mass estimates and
      also in a systematic bias with respect to the true mass. 
      The red arrow shows the median error
      for our sample of classical satellites. 
    }
    \label{fig:MWObs_Vary}
\end{figure}

Fig.~\ref{fig:MWObs_Vary} shows the MW halo mass estimate inferred
from the $(E,L)$ method as a function of the size of the proper motion
errors, $\sigma_{\mu}$. As we increase $\sigma_{\mu}$, we find, as
expected, that the uncertainty in the mass determination
increases. However, the current proper motion errors for the classical
satellites are so small that they fall in the region where there is
hardly any dependence of the mass estimate on
$\sigma_{\mu}$. Improving the current observational errors will
provide little improvement on the mass estimate.

More importantly, we also find a systematic shift in the estimated
halo mass, which increases rapidly with the size of the proper motion
errors. For example, for
$\sigma_{\mu} \approx 0.35~\textnormal{mas/year}$, the estimated mass
is a factor of two too high. This comes about because large proper
motion errors bias the observed velocities high, thus leading to
higher energy and angular momentum values, which, in turn, lead to
higher mass estimates. This is not a problem for our current estimate
since all the classical satellites have proper motions errors well
below $0.1~\textnormal{mas/year}$, and thus lie in the region where
the mass estimate is flat. However, were we to include in the sample
ultrafaint dwarf satellites, many of which have large proper motion
errors \citep[e.g.][]{Fritz2018}, then we would need to account for
the additional bias introduced by the observational errors. 

\begin{figure}
    \includegraphics[width=\columnwidth]{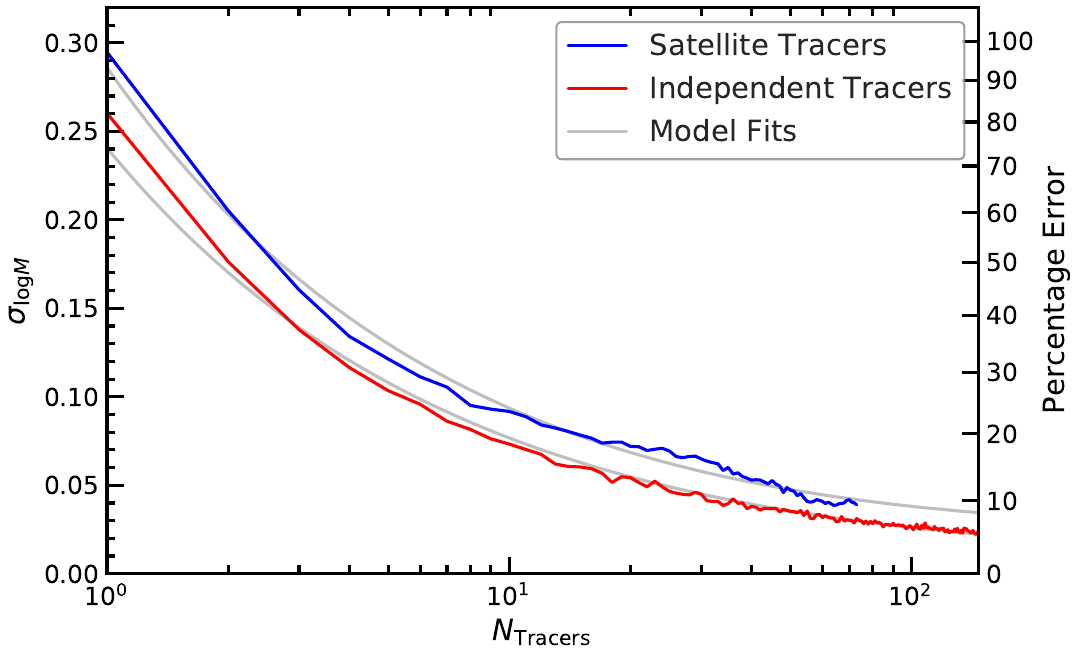}
    \smallspace{}
    \caption{The $1\sigma$ uncertainty, $\sigma_{\log_{10} M}$, with
      which we can determine the logarithm of the halo mass as a
      function of the number of satellite galaxy tracers,
      $N_\textnormal{tracers}$, included in the sample. We show the
      mean uncertainty for a large number of \eagle{} haloes whose
      mass was determined using the $E$ and $L$ values of their most
      massive $N_\textnormal{tracers}$ satellites. The right-hand axis
      shows the percentage errors in $M_{200}$ corresponding to the
      $\sigma_{\log_{10} M}$ values. \change{The blue line gives the results using the satellites of \eagle{} galaxies.
      The red line gives the results from idealised cases of independent satellite tracers (see main text) and represents the statistical limit of our method. The two grey lines show the best fitting curves using Eq. \eqref{eq:sigma_fit}.}}
    \label{fig:NSat_Tracers}
\end{figure}  

The MW is predicted to have approximately 125 satellites brighter than
$M_V=0$, of which just over 50 have already been discovered
\citep{Newton18_MWSatPop_MNRAS.479.2853N}. This means that, in
principle, many more satellites can be used to determine the MW halo
mass, potentially with a smaller uncertainty.
Fig.~\ref{fig:NSat_Tracers} quantifies how the uncertainty in halo
mass is reduced as the number of satellite galaxies in the sample
increases. Here, we consider the simplified case where there are no
observational errors and focus only on the variation arising from the
number of tracers, $N_\textnormal{Tracers}$. 

\change{
Using the same sample of \eagle{} main
haloes as in Section~\ref{Sec:EagleSims}, we determine the host halo mass
using the most massive $N_\textnormal{Tracers}$ subhaloes. To obtain large enough
tracer counts in \eagle{}, we relax the criteria and  consider not only luminous satellites,
but also dark subhaloes. Many of these would be the hosts of the
ultrafaint dwarfs, but \eagle{} lacks the resolution to populate them
with stars. However, these dark substructures are well resolved and
their orbital properties are reliable.  To estimate
an average error for each value of $N_\textnormal{Tracers}$, we
calculate the dispersion in the distribution of
$\log_{10}(M^\textnormal{Esti}_{200}/M^\textnormal{True}_{200})$: the
logarithm of the ratio of estimated to true mass. To ensure accurate measures of the average error, we require at least 100 systems that have $N_\textnormal{Tracers}$ or more tracers; this limits our analysis to $N_\textnormal{tracers} \leq 72$.} 

\change{
The blue line in Fig.~\ref{fig:NSat_Tracers} shows that the expected error in our mass estimate, $\sigma_{\log_{10} M}$, decreases  as the number of tracers increases. We would expect that above a certain number of tracers, the mass determination does not improve any more because the error becomes dominated by halo-to-halo variation and systematic effects such as correlations between the kinematics of different satellites \citep[see e.g.][]{Wang2017,Wang2018}. 
}

\change{
To investigate these effects, we  construct idealised systems by selecting $N_\textnormal{Tracers}$  satellites from our samples' DF, \FEL{}, and then scale them to the mass of random host haloes selected from our sample. This gives us a population of systems whose satellites are perfectly described as being independently drawn from our distribution. As an additional advantage, we are not limited to $N_\textnormal{Tracers} \leq 72$, and can continue increasing $N_\textnormal{Tracers}$ as $\sigma_{\log_{10} M}$ asymptotes to zero (Fig.~\ref{fig:NSat_Tracers}, red line). The difference between the errors in the two samples is the error due to halo-to-halo scatter, $\sigma_\textnormal{Scatter}$. The dependence of the total error, $\sigma_{\log_{10} M}$, on $N_\textnormal{Tracers}$ can be modelled as (cf. \citealt{Li2017ApJ...850..116L}):
\begin{equation}
    \sigma^{2}_{\log_{10} M}=\dfrac{\sigma^{2}_\textnormal{Stat}}{N_\textnormal{Tracers}}+\sigma^{2}_\textnormal{Scatter}
    \label{eq:sigma_fit} \;.
\end{equation}
}

\change{
The mass error for the true \eagle{} satellite systems is best fitted by $\sigma_\textnormal{Stat}=0.29$  and $\sigma_\textnormal{Scatter}=0.03$, while the error for the idealized systems of independent tracers is best described by $\sigma_\textnormal{Stat}=0.24$  and $\sigma_\textnormal{Scatter}=0.01$. We note that a scatter error, $\sigma_\textnormal{Scatter}=0.03$,  equates to an accuracy limit of around $5\%$ and would represent the best mass measurement of the method in the limit of a very large number of tracers. For $10$ satellite tracers we obtain a $\sim 20\%$ uncertainty, similar to our MW mass estimate, while the idealised mass estimates give a slightly smaller uncertainty of $\sim 16\%$. The fits suggest that a $\sim 10\%$ determination of the MW mass is achievable by applying our method to around $N_\textnormal{tracers}\approx 60$ tracers. 
The accuracy of our halo mass measurement could be further improved by
considering the dependence of the satellite dynamics on the properties
and assembly history of the host halo. It is conceivable that by
restricting the analysis to a subset of haloes that more closely
resembles the MW, such as haloes with a similar assembly history, the
halo-to-halo variation could be reduced, leading to an even more
precise halo mass determination. However, at present, the largest benefit would accrue from increasing the number of tracers.
}
\section{Conclusions}
\label{sec:conclusion}

We have developed a method to determine the total mass of the Milky
Way (MW) dark matter (DM) halo by comparing the energy and angular
momentum of MW satellites with the respective distributions predicted
in the \eagle{} galaxy formation cosmological simulations. When scaled
appropriately by host halo mass, the energy and angular momentum of
the satellites become independent of the host halo mass (see
Fig.~\ref{fig:appendix_scaled_E_L}). Thus, we can use a large sample
of \eagle{} haloes, and associated satellites, in our estimate of the
MW halo mass. For this, we constructed the satellite distribution
function in $(E,L)$ space from the simulations and carried out a
maximum likelihood analysis to infer the halo mass from the
phase-space properties of the ten brightest satellite galaxies
(excluding the disrupting Sagittarius galaxy). Using mock samples from
\eagle{} we analysed the performance of the method and quantified its
statistical and systematic uncertainties.

A key test of our method was to apply it to estimate the masses of the
DM haloes of 30 MW analogues simulated in the \auriga{} project. These
simulations have much higher resolution and employ different baryonic
physics models than \eagle. They produce realistic MW-like galaxies
\citep{Auriga_2017MNRAS.467..179G,Grand2018} and thus provide a rigorous and completely
independent external test of our method. We find that our method
provides an unbiased estimate of the total halo masses of the
\auriga{} galaxies, with a precision of $\sim 16\%$, in very good
agreement with the expectations from the \eagle{} simulations.

Our main conclusions are:
\begin{itemize}

\item Applying our method to ten classical MW satellites gives an
  estimate for the total mass of the MW halo of
  $M_{200}^{\textnormal{MW}}=1.17_{-0.15}^{+0.21}\times10^{12}M_{\odot}$. This
  result agrees well with most previous estimates in the literature
  but with a rigorously tested accuracy (${\sim}15\%$) which is better than most other
  estimates.

\item \change{Combining our total DM halo mass estimate with recent estimates
  of the halo mass within $20\kpc{}$ gives an inner DM mass fraction, $M^\textnormal{DM}(<20\kpc)/M^\textnormal{DM}_{200} \approx 0.12$. Assuming that the MW halo follows an NFW profile, we have inferred a Galactic concentration, $C =10.9^{+2.6}_{-2.0}$. 
  This is higher than typical \eagle{} haloes with masses of $10^{12}M_\odot$, which have a median concentration of 8.2, with only ${\sim}23\%$ of them having concentrations of 10.9 or higher. The discrepancy likely reflects that an NFW profile is not a good description of the inner region since the Galactic halo has contracted due to the baryonic components. In fact, when comparing the inner DM mass fraction of the MW against the \eagle{} and \auriga{} simulations, our galaxy is typical of similar mass haloes.}

\item Our halo mass estimate can be improved by increasing the number
  of halo tracers and/or reducing the observational uncertainties. We
  found that the observed proper motions of the ten classical
  satellites are already so precise that further improvement will make
  little difference to the halo mass estimate. Increasing the number
  of satellites, on the other hand, for example by including the
  ${\sim}50$ currently known satellites in the MW, would reduce the
  mass errors to ${\sim}11\%$. Further improvements would be possible
  by analysing all ${\sim}125$ satellites that are predicted to reside
  in the MW \citep{Newton18_MWSatPop_MNRAS.479.2853N}, which would
  result in a ${\sim}8\%$ mass uncertainty, a factor of two
  improvement over our current estimate. 
\end{itemize} 
\noindent

In summary, our MW halo mass estimate is precise and accurate and has
been thoroughly tested on realistic model galaxies and their satellite
populations. Mass estimates that rely on cosmological simulations are
relatively new but the use of simulations enables a robust and
testable methodology. Indeed, the accuracy we are now able to achieve
($\sim 15-20\%$; see also \citealt{PatelMWEnsemble2018ApJ...857...78P})
is a significant step forward from the factor of two uncertainty that
has plagued MW mass estimates for years. This theoretical boost,
coupled with the exquisite 6 dimensional data that \textit{Gaia} and
complementary facilities are now providing, brings us closer to what
may be called the era of ``precision'' near-field cosmology --- when
we can go beyond rough estimates of the MW halo mass and, instead,
remove this important degree of freedom when making use of the
properties of the MW to inform cosmological models and dark matter
theories.

\section*{Acknowledgements}

\change{We thank the anonymous referee for their insightful comments. We also thank Douglas Bourbert, Azadeh Fattahi and Andrew Robertson for helpful comments and discussions.}
TC, MC and CSF were supported by the Science and Technology Facilities
Council (STFC) [grant number ST/F001166/1, ST/I00162X/1,
ST/P000541/1]. AD is supported by a Royal Society University Research
Fellowship. CSF acknowledges European Research Council (ERC) Advanced
Investigator grant DMIDAS (GA 786910). This work used the DiRAC Data
Centric system at Durham University, operated by ICC on behalf of the
STFC DiRAC HPC Facility (www.dirac.ac.uk). This equipment was funded
by BIS National E-infrastructure capital grant ST/K00042X/1, STFC
capital grant ST/H008519/1, and STFC DiRAC Operations grant
ST/K003267/1 and Durham University. DiRAC is part of the National
E-Infrastructure.



\bibliographystyle{mnras}
\bibliography{MWmassBib} 

\begin{thebibliography}{}
\makeatletter
\relax
\def\mn@urlcharsother{\let\do\@makeother \do\$\do\&\do\#\do\^\do\_\do\%\do\~}
\def\mn@doi{\begingroup\mn@urlcharsother \@ifnextchar [ {\mn@doi@}
  {\mn@doi@[]}}
\def\mn@doi@[#1]#2{\def\@tempa{#1}\ifx\@tempa\@empty \href
  {http://dx.doi.org/#2} {doi:#2}\else \href {http://dx.doi.org/#2} {#1}\fi
  \endgroup}
\def\mn@eprint#1#2{\mn@eprint@#1:#2::\@nil}
\def\mn@eprint@arXiv#1{\href {http://arxiv.org/abs/#1} {{\tt arXiv:#1}}}
\def\mn@eprint@dblp#1{\href {http://dblp.uni-trier.de/rec/bibtex/#1.xml}
  {dblp:#1}}
\def\mn@eprint@#1:#2:#3:#4\@nil{\def\@tempa {#1}\def\@tempb {#2}\def\@tempc
  {#3}\ifx \@tempc \@empty \let \@tempc \@tempb \let \@tempb \@tempa \fi \ifx
  \@tempb \@empty \def\@tempb {arXiv}\fi \@ifundefined
  {mn@eprint@\@tempb}{\@tempb:\@tempc}{\expandafter \expandafter \csname
  mn@eprint@\@tempb\endcsname \expandafter{\@tempc}}}

\bibitem[\protect\citeauthoryear{{Barber}, {Starkenburg}, {Navarro},
  {McConnachie}  \& {Fattahi}}{{Barber} et~al.}{2014}]{Barber2014}
{Barber} C.,  {Starkenburg} E.,  {Navarro} J.~F.,  {McConnachie} A.~W.,
  {Fattahi} A.,  2014, \mn@doi [\mnras] {10.1093/mnras/stt1959}, \href
  {http://adsabs.harvard.edu/abs/2014MNRAS.437..959B} {437, 959}

\bibitem[\protect\citeauthoryear{{Binney} \& {Wong}}{{Binney} \&
  {Wong}}{2017}]{BinneyWong2017}
{Binney} J.,  {Wong} L.~K.,  2017, \mn@doi [\mnras] {10.1093/mnras/stx234},
  \href {http://adsabs.harvard.edu/abs/2017MNRAS.467.2446B} {467, 2446}

\bibitem[\protect\citeauthoryear{{Bose}, {Deason}  \& {Frenk}}{{Bose}
  et~al.}{2018}]{Bose2018}
{Bose} S.,  {Deason} A.~J.,   {Frenk} C.~S.,  2018, \mn@doi [\apj]
  {10.3847/1538-4357/aacbc4}, \href
  {http://adsabs.harvard.edu/abs/2018ApJ...863..123B} {863, 123}

\bibitem[\protect\citeauthoryear{{Bowden}, {Belokurov}  \& {Evans}}{{Bowden}
  et~al.}{2015}]{Bowden2015}
{Bowden} A.,  {Belokurov} V.,   {Evans} N.~W.,  2015, \mn@doi [\mnras]
  {10.1093/mnras/stv285}, \href
  {http://adsabs.harvard.edu/abs/2015MNRAS.449.1391B} {449, 1391}

\bibitem[\protect\citeauthoryear{{Bower}, {Schaye}, {Frenk}, {Theuns},
  {Schaller}, {Crain}  \& {McAlpine}}{{Bower} et~al.}{2017}]{Bower2017}
{Bower} R.~G.,  {Schaye} J.,  {Frenk} C.~S.,  {Theuns} T.,  {Schaller} M.,
  {Crain} R.~A.,   {McAlpine} S.,  2017, \mn@doi [\mnras]
  {10.1093/mnras/stw2735}, \href
  {http://adsabs.harvard.edu/abs/2017MNRAS.465...32B} {465, 32}

\bibitem[\protect\citeauthoryear{{Boylan-Kolchin}, {Bullock}  \&
  {Kaplinghat}}{{Boylan-Kolchin} et~al.}{2011}]{BK11_TBtF_MNRAS.415L..40B}
{Boylan-Kolchin} M.,  {Bullock} J.~S.,   {Kaplinghat} M.,  2011, \mn@doi
  [\mnras] {10.1111/j.1745-3933.2011.01074.x}, \href
  {http://adsabs.harvard.edu/abs/2011MNRAS.415L..40B} {415, L40}

\bibitem[\protect\citeauthoryear{{Boylan-Kolchin}, {Bullock}, {Sohn}, {Besla}
  \& {van der Marel}}{{Boylan-Kolchin} et~al.}{2013}]{BK13ApJ...768..140B}
{Boylan-Kolchin} M.,  {Bullock} J.~S.,  {Sohn} S.~T.,  {Besla} G.,   {van der
  Marel} R.~P.,  2013, \mn@doi [\apj] {10.1088/0004-637X/768/2/140}, \href
  {http://adsabs.harvard.edu/abs/2013ApJ...768..140B} {768, 140}

\bibitem[\protect\citeauthoryear{{Busha}, {Marshall}, {Wechsler}, {Klypin}  \&
  {Primack}}{{Busha} et~al.}{2011a}]{Busha2011ApJ...743...40B}
{Busha} M.~T.,  {Marshall} P.~J.,  {Wechsler} R.~H.,  {Klypin} A.,   {Primack}
  J.,  2011a, \mn@doi [\apj] {10.1088/0004-637X/743/1/40}, \href
  {http://adsabs.harvard.edu/abs/2011ApJ...743...40B} {743, 40}

\bibitem[\protect\citeauthoryear{{Busha}, {Wechsler}, {Behroozi}, {Gerke},
  {Klypin}  \& {Primack}}{{Busha} et~al.}{2011b}]{Busha2011b}
{Busha} M.~T.,  {Wechsler} R.~H.,  {Behroozi} P.~S.,  {Gerke} B.~F.,  {Klypin}
  A.~A.,   {Primack} J.~R.,  2011b, \mn@doi [\apj]
  {10.1088/0004-637X/743/2/117}, \href
  {http://adsabs.harvard.edu/abs/2011ApJ...743..117B} {743, 117}

\bibitem[\protect\citeauthoryear{{Cautun} \& {Frenk}}{{Cautun} \&
  {Frenk}}{2017}]{Cautun2017}
{Cautun} M.,  {Frenk} C.~S.,  2017, \mn@doi [\mnras] {10.1093/mnrasl/slx025},
  \href {http://adsabs.harvard.edu/abs/2017MNRAS.468L..41C} {468, L41}

\bibitem[\protect\citeauthoryear{{Cautun}, {Hellwing}, {van de Weygaert},
  {Frenk}, {Jones}  \& {Sawala}}{{Cautun} et~al.}{2014a}]{Cautun2014b}
{Cautun} M.,  {Hellwing} W.~A.,  {van de Weygaert} R.,  {Frenk} C.~S.,  {Jones}
  B.~J.~T.,   {Sawala} T.,  2014a, \mn@doi [\mnras] {10.1093/mnras/stu1829},
  \href {http://adsabs.harvard.edu/abs/2014MNRAS.445.1820C} {445, 1820}

\bibitem[\protect\citeauthoryear{{Cautun}, {Frenk}, {van de Weygaert},
  {Hellwing}  \& {Jones}}{{Cautun} et~al.}{2014b}]{Cautun2014MNRAS.445.2049C}
{Cautun} M.,  {Frenk} C.~S.,  {van de Weygaert} R.,  {Hellwing} W.~A.,
  {Jones} B.~J.~T.,  2014b, \mn@doi [\mnras] {10.1093/mnras/stu1849}, \href
  {http://adsabs.harvard.edu/abs/2014MNRAS.445.2049C} {445, 2049}

\bibitem[\protect\citeauthoryear{{Cautun}, {Bose}, {Frenk}, {Guo}, {Han},
  {Hellwing}, {Sawala}  \& {Wang}}{{Cautun} et~al.}{2015}]{Cautun2015}
{Cautun} M.,  {Bose} S.,  {Frenk} C.~S.,  {Guo} Q.,  {Han} J.,  {Hellwing}
  W.~A.,  {Sawala} T.,   {Wang} W.,  2015, \mn@doi [\mnras]
  {10.1093/mnras/stv1557}, \href
  {http://adsabs.harvard.edu/abs/2015MNRAS.452.3838C} {452, 3838}

\bibitem[\protect\citeauthoryear{{Cole} \& {Binney}}{{Cole} \&
  {Binney}}{2017}]{ColeBinneyCore2017MNRAS.465..798C}
{Cole} D.~R.,  {Binney} J.,  2017, \mn@doi [\mnras] {10.1093/mnras/stw2775},
  \href {http://adsabs.harvard.edu/abs/2017MNRAS.465..798C} {465, 798}

\bibitem[\protect\citeauthoryear{{Crain} et~al.,}{{Crain}
  et~al.}{2015}]{Crain2015}
{Crain} R.~A.,  et~al., 2015, \mn@doi [\mnras] {10.1093/mnras/stv725}, \href
  {http://adsabs.harvard.edu/abs/2015MNRAS.450.1937C} {450, 1937}

\bibitem[\protect\citeauthoryear{{Deason}, {Belokurov}, {Evans}  \&
  {An}}{{Deason} et~al.}{2012}]{Deason12_BrokenDegenMNRAS.424L..44D}
{Deason} A.~J.,  {Belokurov} V.,  {Evans} N.~W.,   {An} J.,  2012, \mn@doi
  [\mnras] {10.1111/j.1745-3933.2012.01283.x}, \href
  {http://adsabs.harvard.edu/abs/2012MNRAS.424L..44D} {424, L44}

\bibitem[\protect\citeauthoryear{{Deason}, {Wetzel}, {Garrison-Kimmel}  \&
  {Belokurov}}{{Deason} et~al.}{2015}]{Deason2015}
{Deason} A.~J.,  {Wetzel} A.~R.,  {Garrison-Kimmel} S.,   {Belokurov} V.,
  2015, \mn@doi [\mnras] {10.1093/mnras/stv1939}, \href
  {http://adsabs.harvard.edu/abs/2015MNRAS.453.3568D} {453, 3568}

\bibitem[\protect\citeauthoryear{{Eadie} \& {Harris}}{{Eadie} \&
  {Harris}}{2016}]{Eadie2016}
{Eadie} G.~M.,  {Harris} W.~E.,  2016, \mn@doi [\apj]
  {10.3847/0004-637X/829/2/108}, \href
  {http://adsabs.harvard.edu/abs/2016ApJ...829..108E} {829, 108}

\bibitem[\protect\citeauthoryear{{Eadie}, {Harris}  \& {Widrow}}{{Eadie}
  et~al.}{2015}]{Eadie2015}
{Eadie} G.~M.,  {Harris} W.~E.,   {Widrow} L.~M.,  2015, \mn@doi [\apj]
  {10.1088/0004-637X/806/1/54}, \href
  {http://adsabs.harvard.edu/abs/2015ApJ...806...54E} {806, 54}

\bibitem[\protect\citeauthoryear{{Evans}, {Wilkinson}, {Perrett}  \&
  {Bridges}}{{Evans} et~al.}{2003}]{Evans2003}
{Evans} N.~W.,  {Wilkinson} M.~I.,  {Perrett} K.~M.,   {Bridges} T.~J.,  2003,
  \mn@doi [\apj] {10.1086/345400}, \href
  {http://adsabs.harvard.edu/abs/2003ApJ...583..752E} {583, 752}

\bibitem[\protect\citeauthoryear{{Fragione} \& {Loeb}}{{Fragione} \&
  {Loeb}}{2017}]{Fragione17_HVSMWMassNewA...55...32F}
{Fragione} G.,  {Loeb} A.,  2017, \mn@doi [\na] {10.1016/j.newast.2017.03.002},
  \href {http://adsabs.harvard.edu/abs/2017NewA...55...32F} {55, 32}

\bibitem[\protect\citeauthoryear{{Fritz}, {Battaglia}, {Pawlowski},
  {Kallivayalil}, {van der Marel}, {Sohn}, {Brook}  \& {Besla}}{{Fritz}
  et~al.}{2018}]{Fritz2018}
{Fritz} T.~K.,  {Battaglia} G.,  {Pawlowski} M.~S.,  {Kallivayalil} N.,  {van
  der Marel} R.,  {Sohn} S.~T.,  {Brook} C.,   {Besla} G.,  2018, \mn@doi
  [\aap] {10.1051/0004-6361/201833343}, \href
  {http://adsabs.harvard.edu/abs/2018A%26A...619A.103F} {619, A103}

\bibitem[\protect\citeauthoryear{{Gaia Collaboration} et~al.,}{{Gaia
  Collaboration} et~al.}{2018a}]{GaiaDR22018arXiv180409365G}
{Gaia Collaboration} et~al., 2018a, \mn@doi [\aap]
  {10.1051/0004-6361/201833051}, \href
  {http://adsabs.harvard.edu/abs/2018A%26A...616A...1G} {616, A1}

\bibitem[\protect\citeauthoryear{{Gaia Collaboration} et~al.,}{{Gaia
  Collaboration} et~al.}{2018b}]{GaiaDR2_MotionsDwarfGC_2018arXiv180409381G}
{Gaia Collaboration} et~al., 2018b, \mn@doi [\aap]
  {10.1051/0004-6361/201832698}, \href
  {http://adsabs.harvard.edu/abs/2018A%26A...616A..12G} {616, A12}

\bibitem[\protect\citeauthoryear{{Gibbons}, {Belokurov}  \& {Evans}}{{Gibbons}
  et~al.}{2014}]{Gibbons2014}
{Gibbons} S.~L.~J.,  {Belokurov} V.,   {Evans} N.~W.,  2014, \mn@doi [\mnras]
  {10.1093/mnras/stu1986}, \href
  {http://adsabs.harvard.edu/abs/2014MNRAS.445.3788G} {445, 3788}

\bibitem[\protect\citeauthoryear{{Gnedin}, {Kravtsov}, {Klypin}  \&
  {Nagai}}{{Gnedin} et~al.}{2004}]{Gnedin2004}
{Gnedin} O.~Y.,  {Kravtsov} A.~V.,  {Klypin} A.~A.,   {Nagai} D.,  2004,
  \mn@doi [\apj] {10.1086/424914}, \href
  {http://adsabs.harvard.edu/abs/2004ApJ...616...16G} {616, 16}

\bibitem[\protect\citeauthoryear{{Gnedin}, {Brown}, {Geller}  \&
  {Kenyon}}{{Gnedin} et~al.}{2010}]{Gnedin2010}
{Gnedin} O.~Y.,  {Brown} W.~R.,  {Geller} M.~J.,   {Kenyon} S.~J.,  2010,
  \mn@doi [\apjl] {10.1088/2041-8205/720/1/L108}, \href
  {http://adsabs.harvard.edu/abs/2010ApJ...720L.108G} {720, L108}

\bibitem[\protect\citeauthoryear{{G{\'o}mez}, {Besla}, {Carpintero},
  {Villalobos}, {O'Shea}  \& {Bell}}{{G{\'o}mez} et~al.}{2015}]{Gomez2015}
{G{\'o}mez} F.~A.,  {Besla} G.,  {Carpintero} D.~D.,  {Villalobos} {\'A}.,
  {O'Shea} B.~W.,   {Bell} E.~F.,  2015, \mn@doi [\apj]
  {10.1088/0004-637X/802/2/128}, \href
  {http://adsabs.harvard.edu/abs/2015ApJ...802..128G} {802, 128}

\bibitem[\protect\citeauthoryear{{Gonz{\'a}lez}, {Kravtsov}  \&
  {Gnedin}}{{Gonz{\'a}lez} et~al.}{2013}]{Gonzalez2013}
{Gonz{\'a}lez} R.~E.,  {Kravtsov} A.~V.,   {Gnedin} N.~Y.,  2013, \mn@doi
  [\apj] {10.1088/0004-637X/770/2/96}, \href
  {http://adsabs.harvard.edu/abs/2013ApJ...770...96G} {770, 96}

\bibitem[\protect\citeauthoryear{{Grand} et~al.,}{{Grand}
  et~al.}{2017}]{Auriga_2017MNRAS.467..179G}
{Grand} R.~J.~J.,  et~al., 2017, \mn@doi [\mnras] {10.1093/mnras/stx071}, \href
  {http://adsabs.harvard.edu/abs/2017MNRAS.467..179G} {467, 179}

\bibitem[\protect\citeauthoryear{{Grand} et~al.,}{{Grand}
  et~al.}{2018}]{Grand2018}
{Grand} R.~J.~J.,  et~al., 2018, \mn@doi [\mnras] {10.1093/mnras/sty2403},
  \href {http://adsabs.harvard.edu/abs/2018MNRAS.481.1726G} {481, 1726}

\bibitem[\protect\citeauthoryear{{Han}, {Wang}, {Cole}  \& {Frenk}}{{Han}
  et~al.}{2016a}]{Han2016a}
{Han} J.,  {Wang} W.,  {Cole} S.,   {Frenk} C.~S.,  2016a, \mn@doi [\mnras]
  {10.1093/mnras/stv2707}, \href
  {http://adsabs.harvard.edu/abs/2016MNRAS.456.1003H} {456, 1003}

\bibitem[\protect\citeauthoryear{{Han}, {Wang}, {Cole}  \& {Frenk}}{{Han}
  et~al.}{2016b}]{Han2016b}
{Han} J.,  {Wang} W.,  {Cole} S.,   {Frenk} C.~S.,  2016b, \mn@doi [\mnras]
  {10.1093/mnras/stv2522}, \href
  {http://adsabs.harvard.edu/abs/2016MNRAS.456.1017H} {456, 1017}

\bibitem[\protect\citeauthoryear{{Hellwing}, {Frenk}, {Cautun}, {Bose},
  {Helly}, {Jenkins}, {Sawala}  \& {Cytowski}}{{Hellwing}
  et~al.}{2016}]{Hellwing2016}
{Hellwing} W.~A.,  {Frenk} C.~S.,  {Cautun} M.,  {Bose} S.,  {Helly} J.,
  {Jenkins} A.,  {Sawala} T.,   {Cytowski} M.,  2016, \mn@doi [\mnras]
  {10.1093/mnras/stw214}, \href
  {http://adsabs.harvard.edu/abs/2016MNRAS.457.3492H} {457, 3492}

\bibitem[\protect\citeauthoryear{{Huang} et~al.,}{{Huang}
  et~al.}{2016}]{Huang2016}
{Huang} Y.,  et~al., 2016, \mn@doi [\mnras] {10.1093/mnras/stw2096}, \href
  {http://adsabs.harvard.edu/abs/2016MNRAS.463.2623H} {463, 2623}

\bibitem[\protect\citeauthoryear{{Kafle}, {Sharma}, {Lewis}  \&
  {Bland-Hawthorn}}{{Kafle} et~al.}{2012}]{Kafle2012}
{Kafle} P.~R.,  {Sharma} S.,  {Lewis} G.~F.,   {Bland-Hawthorn} J.,  2012,
  \mn@doi [\apj] {10.1088/0004-637X/761/2/98}, \href
  {http://adsabs.harvard.edu/abs/2012ApJ...761...98K} {761, 98}

\bibitem[\protect\citeauthoryear{{Kafle}, {Sharma}, {Lewis}  \&
  {Bland-Hawthorn}}{{Kafle} et~al.}{2014}]{Kafle2014}
{Kafle} P.~R.,  {Sharma} S.,  {Lewis} G.~F.,   {Bland-Hawthorn} J.,  2014,
  \mn@doi [\apj] {10.1088/0004-637X/794/1/59}, \href
  {http://adsabs.harvard.edu/abs/2014ApJ...794...59K} {794, 59}

\bibitem[\protect\citeauthoryear{{Kallivayalil}, {van der Marel}, {Besla},
  {Anderson}  \& {Alcock}}{{Kallivayalil} et~al.}{2013}]{Kallivayalil2013}
{Kallivayalil} N.,  {van der Marel} R.~P.,  {Besla} G.,  {Anderson} J.,
  {Alcock} C.,  2013, \mn@doi [\apj] {10.1088/0004-637X/764/2/161}, \href
  {http://adsabs.harvard.edu/abs/2013ApJ...764..161K} {764, 161}

\bibitem[\protect\citeauthoryear{{Kennedy}, {Frenk}, {Cole}  \&
  {Benson}}{{Kennedy}
  et~al.}{2014}]{Kennedy2014:WDMParticlefromMWSat_MNRAS.442.2487K}
{Kennedy} R.,  {Frenk} C.,  {Cole} S.,   {Benson} A.,  2014, \mn@doi [\mnras]
  {10.1093/mnras/stu719}, \href
  {http://adsabs.harvard.edu/abs/2014MNRAS.442.2487K} {442, 2487}

\bibitem[\protect\citeauthoryear{{Klypin}, {Kravtsov}, {Valenzuela}  \&
  {Prada}}{{Klypin} et~al.}{1999}]{Klypin1999}
{Klypin} A.,  {Kravtsov} A.~V.,  {Valenzuela} O.,   {Prada} F.,  1999, \mn@doi
  [\apj] {10.1086/307643}, \href
  {http://adsabs.harvard.edu/abs/1999ApJ...522...82K} {522, 82}

\bibitem[\protect\citeauthoryear{{Koposov}, {Rix}  \& {Hogg}}{{Koposov}
  et~al.}{2010}]{Koposov2010}
{Koposov} S.~E.,  {Rix} H.-W.,   {Hogg} D.~W.,  2010, \mn@doi [\apj]
  {10.1088/0004-637X/712/1/260}, \href
  {http://adsabs.harvard.edu/abs/2010ApJ...712..260K} {712, 260}

\bibitem[\protect\citeauthoryear{{K{\"u}pper}, {Balbinot}, {Bonaca},
  {Johnston}, {Hogg}, {Kroupa}  \& {Santiago}}{{K{\"u}pper}
  et~al.}{2015}]{Kupper2015}
{K{\"u}pper} A.~H.~W.,  {Balbinot} E.,  {Bonaca} A.,  {Johnston} K.~V.,  {Hogg}
  D.~W.,  {Kroupa} P.,   {Santiago} B.~X.,  2015, \mn@doi [\apj]
  {10.1088/0004-637X/803/2/80}, \href
  {http://adsabs.harvard.edu/abs/2015ApJ...803...80K} {803, 80}

\bibitem[\protect\citeauthoryear{{Li} \& {White}}{{Li} \&
  {White}}{2008}]{Li2008}
{Li} Y.-S.,  {White} S.~D.~M.,  2008, \mn@doi [\mnras]
  {10.1111/j.1365-2966.2007.12748.x}, \href
  {http://adsabs.harvard.edu/abs/2008MNRAS.384.1459L} {384, 1459}

\bibitem[\protect\citeauthoryear{{Li}, {Jing}, {Qian}, {Yuan}  \& {Zhao}}{{Li}
  et~al.}{2017}]{Li2017ApJ...850..116L}
{Li} Z.-Z.,  {Jing} Y.~P.,  {Qian} Y.-Z.,  {Yuan} Z.,   {Zhao} D.-H.,  2017,
  \mn@doi [\apj] {10.3847/1538-4357/aa94c0}, \href
  {http://adsabs.harvard.edu/abs/2017ApJ...850..116L} {850, 116}

\bibitem[\protect\citeauthoryear{{Lovell}, {Frenk}, {Eke}, {Jenkins}, {Gao}  \&
  {Theuns}}{{Lovell} et~al.}{2014}]{Lovell2014:WarmDM_MNRAS.439..300L}
{Lovell} M.~R.,  {Frenk} C.~S.,  {Eke} V.~R.,  {Jenkins} A.,  {Gao} L.,
  {Theuns} T.,  2014, \mn@doi [\mnras] {10.1093/mnras/stt2431}, \href
  {http://adsabs.harvard.edu/abs/2014MNRAS.439..300L} {439, 300}

\bibitem[\protect\citeauthoryear{{Lu}, {Mo}, {Katz}  \& {Weinberg}}{{Lu}
  et~al.}{2006}]{Lu2006}
{Lu} Y.,  {Mo} H.~J.,  {Katz} N.,   {Weinberg} M.~D.,  2006, \mn@doi [\mnras]
  {10.1111/j.1365-2966.2006.10270.x}, \href
  {http://adsabs.harvard.edu/abs/2006MNRAS.368.1931L} {368, 1931}

\bibitem[\protect\citeauthoryear{{Ludlow}, {Navarro}, {Angulo},
  {Boylan-Kolchin}, {Springel}, {Frenk}  \& {White}}{{Ludlow}
  et~al.}{2014}]{Ludlow2014}
{Ludlow} A.~D.,  {Navarro} J.~F.,  {Angulo} R.~E.,  {Boylan-Kolchin} M.,
  {Springel} V.,  {Frenk} C.,   {White} S.~D.~M.,  2014, \mn@doi [\mnras]
  {10.1093/mnras/stu483}, \href
  {http://adsabs.harvard.edu/abs/2014MNRAS.441..378L} {441, 378}

\bibitem[\protect\citeauthoryear{{Matthee}, {Schaye}, {Crain}, {Schaller},
  {Bower}  \& {Theuns}}{{Matthee} et~al.}{2017}]{Matthee2017}
{Matthee} J.,  {Schaye} J.,  {Crain} R.~A.,  {Schaller} M.,  {Bower} R.,
  {Theuns} T.,  2017, \mn@doi [\mnras] {10.1093/mnras/stw2884}, \href
  {http://adsabs.harvard.edu/abs/2017MNRAS.465.2381M} {465, 2381}

\bibitem[\protect\citeauthoryear{{McConnachie}}{{McConnachie}}{2012}]{Obs_McConnachie12_DwarfPosAJ....144....4M}
{McConnachie} A.~W.,  2012, \mn@doi [\aj] {10.1088/0004-6256/144/1/4}, \href
  {http://adsabs.harvard.edu/abs/2012AJ....144....4M} {144, 4}

\bibitem[\protect\citeauthoryear{{McMillan}}{{McMillan}}{2011}]{McMillan2011}
{McMillan} P.~J.,  2011, \mn@doi [\mnras] {10.1111/j.1365-2966.2011.18564.x},
  \href {http://adsabs.harvard.edu/abs/2011MNRAS.414.2446M} {414, 2446}

\bibitem[\protect\citeauthoryear{{McMillan}}{{McMillan}}{2017}]{McMillan2017}
{McMillan} P.~J.,  2017, \mn@doi [\mnras] {10.1093/mnras/stw2759}, \href
  {http://adsabs.harvard.edu/abs/2017MNRAS.465...76M} {465, 76}

\bibitem[\protect\citeauthoryear{{Monari} et~al.,}{{Monari}
  et~al.}{2018}]{Monari2018}
{Monari} G.,  et~al., 2018, \mn@doi [\aap] {10.1051/0004-6361/201833748}, \href
  {http://adsabs.harvard.edu/abs/2018A%26A...616L...9M} {616, L9}

\bibitem[\protect\citeauthoryear{{Moore}, {Ghigna}, {Governato}, {Lake},
  {Quinn}, {Stadel}  \& {Tozzi}}{{Moore} et~al.}{1999}]{Moore1999}
{Moore} B.,  {Ghigna} S.,  {Governato} F.,  {Lake} G.,  {Quinn} T.,  {Stadel}
  J.,   {Tozzi} P.,  1999, \mn@doi [\apjl] {10.1086/312287}, \href
  {http://adsabs.harvard.edu/abs/1999ApJ...524L..19M} {524, L19}

\bibitem[\protect\citeauthoryear{{Navarro}, {Frenk}  \& {White}}{{Navarro}
  et~al.}{1996}]{Navarro1996}
{Navarro} J.~F.,  {Frenk} C.~S.,   {White} S.~D.~M.,  1996, \mn@doi [\apj]
  {10.1086/177173}, \href {http://adsabs.harvard.edu/abs/1996ApJ...462..563N}
  {462, 563}

\bibitem[\protect\citeauthoryear{{Navarro}, {Frenk}  \& {White}}{{Navarro}
  et~al.}{1997}]{Navarro1997}
{Navarro} J.~F.,  {Frenk} C.~S.,   {White} S.~D.~M.,  1997, \mn@doi [\apj]
  {10.1086/304888}, \href {http://adsabs.harvard.edu/abs/1997ApJ...490..493N}
  {490, 493}

\bibitem[\protect\citeauthoryear{{Neto} et~al.,}{{Neto}
  et~al.}{2007}]{Neto2007}
{Neto} A.~F.,  et~al., 2007, \mn@doi [\mnras]
  {10.1111/j.1365-2966.2007.12381.x}, \href
  {http://adsabs.harvard.edu/abs/2007MNRAS.381.1450N} {381, 1450}

\bibitem[\protect\citeauthoryear{{Newberg}, {Willett}, {Yanny}  \&
  {Xu}}{{Newberg} et~al.}{2010}]{Newberg10_StreamMWMass_ApJ...711...32N}
{Newberg} H.~J.,  {Willett} B.~A.,  {Yanny} B.,   {Xu} Y.,  2010, \mn@doi
  [\apj] {10.1088/0004-637X/711/1/32}, \href
  {http://adsabs.harvard.edu/abs/2010ApJ...711...32N} {711, 32}

\bibitem[\protect\citeauthoryear{{Newton}, {Cautun}, {Jenkins}, {Frenk}  \&
  {Helly}}{{Newton} et~al.}{2018}]{Newton18_MWSatPop_MNRAS.479.2853N}
{Newton} O.,  {Cautun} M.,  {Jenkins} A.,  {Frenk} C.~S.,   {Helly} J.~C.,
  2018, \mn@doi [\mnras] {10.1093/mnras/sty1085}, \href
  {http://adsabs.harvard.edu/abs/2018MNRAS.479.2853N} {479, 2853}

\bibitem[\protect\citeauthoryear{{Patel}, {Besla}  \& {Mandel}}{{Patel}
  et~al.}{2017}]{Patel2017Orbits2_MNRAS.468.3428P}
{Patel} E.,  {Besla} G.,   {Mandel} K.,  2017, \mn@doi [\mnras]
  {10.1093/mnras/stx698}, \href
  {http://adsabs.harvard.edu/abs/2017MNRAS.468.3428P} {468, 3428}

\bibitem[\protect\citeauthoryear{{Patel}, {Besla}, {Mandel}  \& {Sohn}}{{Patel}
  et~al.}{2018}]{PatelMWEnsemble2018ApJ...857...78P}
{Patel} E.,  {Besla} G.,  {Mandel} K.,   {Sohn} S.~T.,  2018, \mn@doi [\apj]
  {10.3847/1538-4357/aab78f}, \href
  {http://adsabs.harvard.edu/abs/2018ApJ...857...78P} {857, 78}

\bibitem[\protect\citeauthoryear{{Pawlowski} et~al.,}{{Pawlowski}
  et~al.}{2014}]{Pawlowski2014}
{Pawlowski} M.~S.,  et~al., 2014, \mn@doi [\mnras] {10.1093/mnras/stu1005},
  \href {http://adsabs.harvard.edu/abs/2014MNRAS.442.2362P} {442, 2362}

\bibitem[\protect\citeauthoryear{{Pe{\~n}arrubia} \&
  {Fattahi}}{{Pe{\~n}arrubia} \& {Fattahi}}{2017}]{Penarrubia2017}
{Pe{\~n}arrubia} J.,  {Fattahi} A.,  2017, \mn@doi [\mnras]
  {10.1093/mnras/stx323}, \href
  {http://adsabs.harvard.edu/abs/2017MNRAS.468.1300P} {468, 1300}

\bibitem[\protect\citeauthoryear{{Pe{\~n}arrubia}, {G{\'o}mez}, {Besla},
  {Erkal}  \& {Ma}}{{Pe{\~n}arrubia} et~al.}{2016}]{Penarrubia2016}
{Pe{\~n}arrubia} J.,  {G{\'o}mez} F.~A.,  {Besla} G.,  {Erkal} D.,   {Ma}
  Y.-Z.,  2016, \mn@doi [\mnras] {10.1093/mnrasl/slv160}, \href
  {http://adsabs.harvard.edu/abs/2016MNRAS.456L..54P} {456, L54}

\bibitem[\protect\citeauthoryear{{Pfeffer}, {Kruijssen}, {Crain}  \&
  {Bastian}}{{Pfeffer} et~al.}{2018}]{Pfeffer2018}
{Pfeffer} J.,  {Kruijssen} J.~M.~D.,  {Crain} R.~A.,   {Bastian} N.,  2018,
  \mn@doi [\mnras] {10.1093/mnras/stx3124}, \href
  {http://adsabs.harvard.edu/abs/2018MNRAS.475.4309P} {475, 4309}

\bibitem[\protect\citeauthoryear{{Piatek}, {Pryor}  \& {Olszewski}}{{Piatek}
  et~al.}{2016}]{Obs_PMLeoII_2016AJ....152..166P}
{Piatek} S.,  {Pryor} C.,   {Olszewski} E.~W.,  2016, \mn@doi [\aj]
  {10.3847/0004-6256/152/6/166}, \href
  {http://adsabs.harvard.edu/abs/2016AJ....152..166P} {152, 166}

\bibitem[\protect\citeauthoryear{{Piffl} et~al.,}{{Piffl}
  et~al.}{2014}]{Piffl2014}
{Piffl} T.,  et~al., 2014, \mn@doi [\aap] {10.1051/0004-6361/201322531}, \href
  {http://adsabs.harvard.edu/abs/2014A%26A...562A..91P} {562, A91}

\bibitem[\protect\citeauthoryear{{Planck Collaboration} et~al.,}{{Planck
  Collaboration} et~al.}{2014}]{planck2014}
{Planck Collaboration} et~al., 2014, \mn@doi [\aap]
  {10.1051/0004-6361/201321529}, \href
  {http://adsabs.harvard.edu/abs/2014A%26A...571A...1P} {571, A1}

\bibitem[\protect\citeauthoryear{{Posti} \& {Helmi}}{{Posti} \&
  {Helmi}}{2018}]{PostiHelmi2018arXiv180501408P}
{Posti} L.,  {Helmi} A.,  2018, preprint, \href
  {http://adsabs.harvard.edu/abs/2018arXiv180501408P} {} (\mn@eprint {arXiv}
  {1805.01408})

\bibitem[\protect\citeauthoryear{{Purcell} \& {Zentner}}{{Purcell} \&
  {Zentner}}{2012}]{Purcell2012}
{Purcell} C.~W.,  {Zentner} A.~R.,  2012, \mn@doi [\jcap]
  {10.1088/1475-7516/2012/12/007}, \href
  {http://adsabs.harvard.edu/abs/2012JCAP...12..007P} {12, 007}

\bibitem[\protect\citeauthoryear{{Rossi}, {Marchetti}, {Cacciato}, {Kuiack}  \&
  {Sari}}{{Rossi} et~al.}{2017}]{Rossi2017}
{Rossi} E.~M.,  {Marchetti} T.,  {Cacciato} M.,  {Kuiack} M.,   {Sari} R.,
  2017, \mn@doi [\mnras] {10.1093/mnras/stx098}, \href
  {http://adsabs.harvard.edu/abs/2017MNRAS.467.1844R} {467, 1844}

\bibitem[\protect\citeauthoryear{{Schaller} et~al.,}{{Schaller}
  et~al.}{2015}]{MatthieuNFWConc2015MNRAS.451.1247S}
{Schaller} M.,  et~al., 2015, \mn@doi [\mnras] {10.1093/mnras/stv1067}, \href
  {http://adsabs.harvard.edu/abs/2015MNRAS.451.1247S} {451, 1247}

\bibitem[\protect\citeauthoryear{{Schaye} et~al.,}{{Schaye}
  et~al.}{2015}]{EAGLE2015MNRAS.446..521S}
{Schaye} J.,  et~al., 2015, \mn@doi [\mnras] {10.1093/mnras/stu2058}, \href
  {http://adsabs.harvard.edu/abs/2015MNRAS.446..521S} {446, 521}

\bibitem[\protect\citeauthoryear{{Sch{\"o}nrich}, {Binney}  \&
  {Dehnen}}{{Sch{\"o}nrich} et~al.}{2010}]{Schonrich2010}
{Sch{\"o}nrich} R.,  {Binney} J.,   {Dehnen} W.,  2010, \mn@doi [\mnras]
  {10.1111/j.1365-2966.2010.16253.x}, \href
  {http://adsabs.harvard.edu/abs/2010MNRAS.403.1829S} {403, 1829}

\bibitem[\protect\citeauthoryear{{Shao}, {Cautun}  \& {Frenk}}{{Shao}
  et~al.}{2018a}]{Shao2018c}
{Shao} S.,  {Cautun} M.,   {Frenk} C.~S.,  2018a, in prep

\bibitem[\protect\citeauthoryear{{Shao}, {Cautun}, {Frenk}, {Grand},
  {G{\'o}mez}, {Marinacci}  \& {Simpson}}{{Shao} et~al.}{2018b}]{Shao2018a}
{Shao} S.,  {Cautun} M.,  {Frenk} C.~S.,  {Grand} R.~J.~J.,  {G{\'o}mez} F.~A.,
   {Marinacci} F.,   {Simpson} C.~M.,  2018b, \mn@doi [\mnras]
  {10.1093/mnras/sty343}, \href
  {http://adsabs.harvard.edu/abs/2018MNRAS.476.1796S} {476, 1796}

\bibitem[\protect\citeauthoryear{{Shao}, {Cautun}, {Deason}, {Frenk}  \&
  {Theuns}}{{Shao} et~al.}{2018c}]{Shao2018b}
{Shao} S.,  {Cautun} M.,  {Deason} A.~J.,  {Frenk} C.~S.,   {Theuns} T.,
  2018c, \mn@doi [\mnras] {10.1093/mnras/sty1470}, \href
  {http://adsabs.harvard.edu/abs/2018MNRAS.479..284S} {479, 284}

\bibitem[\protect\citeauthoryear{{Simon}}{{Simon}}{2018}]{Simon2018}
{Simon} J.~D.,  2018, \mn@doi [\apj] {10.3847/1538-4357/aacdfb}, \href
  {http://adsabs.harvard.edu/abs/2018ApJ...863...89S} {863, 89}

\bibitem[\protect\citeauthoryear{{Smith} et~al.,}{{Smith}
  et~al.}{2007}]{Smith2007}
{Smith} M.~C.,  et~al., 2007, \mn@doi [\mnras]
  {10.1111/j.1365-2966.2007.11964.x}, \href
  {http://adsabs.harvard.edu/abs/2007MNRAS.379..755S} {379, 755}

\bibitem[\protect\citeauthoryear{{Sohn}, {Besla}, {van der Marel},
  {Boylan-Kolchin}, {Majewski}  \& {Bullock}}{{Sohn}
  et~al.}{2013}]{Obs_PMLeoI_2013ApJ...768..139S}
{Sohn} S.~T.,  {Besla} G.,  {van der Marel} R.~P.,  {Boylan-Kolchin} M.,
  {Majewski} S.~R.,   {Bullock} J.~S.,  2013, \mn@doi [\apj]
  {10.1088/0004-637X/768/2/139}, \href
  {http://adsabs.harvard.edu/abs/2013ApJ...768..139S} {768, 139}

\bibitem[\protect\citeauthoryear{{Sohn}, {Watkins}, {Fardal}, {van der Marel},
  {Deason}, {Besla}  \& {Bellini}}{{Sohn} et~al.}{2018}]{Sohn2018}
{Sohn} S.~T.,  {Watkins} L.~L.,  {Fardal} M.~A.,  {van der Marel} R.~P.,
  {Deason} A.~J.,  {Besla} G.,   {Bellini} A.,  2018, \mn@doi [\apj]
  {10.3847/1538-4357/aacd0b}, \href
  {http://adsabs.harvard.edu/abs/2018ApJ...862...52S} {862, 52}

\bibitem[\protect\citeauthoryear{{Springel}, {Yoshida}  \& {White}}{{Springel}
  et~al.}{2001}]{Springel2001}
{Springel} V.,  {Yoshida} N.,   {White} S.~D.~M.,  2001, \mn@doi [\na]
  {10.1016/S1384-1076(01)00042-2}, \href
  {http://adsabs.harvard.edu/abs/2001NewA....6...79S} {6, 79}

\bibitem[\protect\citeauthoryear{{Springel} et~al.,}{{Springel}
  et~al.}{2008}]{Springel2008}
{Springel} V.,  et~al., 2008, \mn@doi [\mnras]
  {10.1111/j.1365-2966.2008.14066.x}, \href
  {http://adsabs.harvard.edu/abs/2008MNRAS.391.1685S} {391, 1685}

\bibitem[\protect\citeauthoryear{{Vasiliev}}{{Vasiliev}}{2019}]{Vasiliev18_AGAMAarXiv180208239V}
{Vasiliev} E.,  2019, \mn@doi [\mnras] {10.1093/mnras/sty2672}, \href
  {http://adsabs.harvard.edu/abs/2019MNRAS.482.1525V} {482, 1525}

\bibitem[\protect\citeauthoryear{{Vera-Ciro}, {Helmi}, {Starkenburg}  \&
  {Breddels}}{{Vera-Ciro} et~al.}{2013}]{VeraCiro2013}
{Vera-Ciro} C.~A.,  {Helmi} A.,  {Starkenburg} E.,   {Breddels} M.~A.,  2013,
  \mn@doi [\mnras] {10.1093/mnras/sts148}, \href
  {http://adsabs.harvard.edu/abs/2013MNRAS.428.1696V} {428, 1696}

\bibitem[\protect\citeauthoryear{{Wang}, {Frenk}, {Navarro}, {Gao}  \&
  {Sawala}}{{Wang} et~al.}{2012}]{Wang12_MissingSat_MNRAS.424.2715W}
{Wang} J.,  {Frenk} C.~S.,  {Navarro} J.~F.,  {Gao} L.,   {Sawala} T.,  2012,
  \mn@doi [\mnras] {10.1111/j.1365-2966.2012.21357.x}, \href
  {http://adsabs.harvard.edu/abs/2012MNRAS.424.2715W} {424, 2715}

\bibitem[\protect\citeauthoryear{{Wang}, {Han}, {Cooper}, {Cole}, {Frenk}  \&
  {Lowing}}{{Wang} et~al.}{2015}]{Wang15_DMHaloDynTracers_MNRAS.453..377W}
{Wang} W.,  {Han} J.,  {Cooper} A.~P.,  {Cole} S.,  {Frenk} C.,   {Lowing} B.,
  2015, \mn@doi [\mnras] {10.1093/mnras/stv1647}, \href
  {http://adsabs.harvard.edu/abs/2015MNRAS.453..377W} {453, 377}

\bibitem[\protect\citeauthoryear{{Wang}, {Han}, {Cole}, {Frenk}  \&
  {Sawala}}{{Wang} et~al.}{2017}]{Wang2017}
{Wang} W.,  {Han} J.,  {Cole} S.,  {Frenk} C.,   {Sawala} T.,  2017, \mn@doi
  [\mnras] {10.1093/mnras/stx1334}, \href
  {http://adsabs.harvard.edu/abs/2017MNRAS.470.2351W} {470, 2351}

\bibitem[\protect\citeauthoryear{{Wang}, {Han}, {Cole}, {More}, {Frenk}  \&
  {Schaller}}{{Wang} et~al.}{2018}]{Wang2018}
{Wang} W.,  {Han} J.,  {Cole} S.,  {More} S.,  {Frenk} C.,   {Schaller} M.,
  2018, \mn@doi [\mnras] {10.1093/mnras/sty706}, \href
  {http://adsabs.harvard.edu/abs/2018MNRAS.476.5669W} {476, 5669}

\bibitem[\protect\citeauthoryear{{Watkins}, {Evans}  \& {An}}{{Watkins}
  et~al.}{2010}]{Watkins2010}
{Watkins} L.~L.,  {Evans} N.~W.,   {An} J.~H.,  2010, \mn@doi [\mnras]
  {10.1111/j.1365-2966.2010.16708.x}, \href
  {http://adsabs.harvard.edu/abs/2010MNRAS.406..264W} {406, 264}

\bibitem[\protect\citeauthoryear{{Watkins}, {van der Marel}, {Sohn}  \&
  {Evans}}{{Watkins} et~al.}{2018}]{WatkinsMWGC2018arXiv180411348W}
{Watkins} L.~L.,  {van der Marel} R.~P.,  {Sohn} S.~T.,   {Evans} N.~W.,  2018,
  preprint, \href {http://adsabs.harvard.edu/abs/2018arXiv180411348W} {}
  (\mn@eprint {arXiv} {1804.11348})

\bibitem[\protect\citeauthoryear{{Wechsler}, {Bullock}, {Primack}, {Kravtsov}
  \& {Dekel}}{{Wechsler} et~al.}{2002}]{Wechsler2002}
{Wechsler} R.~H.,  {Bullock} J.~S.,  {Primack} J.~R.,  {Kravtsov} A.~V.,
  {Dekel} A.,  2002, \mn@doi [\apj] {10.1086/338765}, \href
  {http://adsabs.harvard.edu/abs/2002ApJ...568...52W} {568, 52}

\bibitem[\protect\citeauthoryear{{Wetzel}, {Deason}  \&
  {Garrison-Kimmel}}{{Wetzel} et~al.}{2015}]{Wetzel2015}
{Wetzel} A.~R.,  {Deason} A.~J.,   {Garrison-Kimmel} S.,  2015, \mn@doi [\apj]
  {10.1088/0004-637X/807/1/49}, \href
  {http://adsabs.harvard.edu/abs/2015ApJ...807...49W} {807, 49}

\bibitem[\protect\citeauthoryear{{Wilkinson} \& {Evans}}{{Wilkinson} \&
  {Evans}}{1999}]{WilkinsonEvans1999}
{Wilkinson} M.~I.,  {Evans} N.~W.,  1999, \mn@doi [\mnras]
  {10.1046/j.1365-8711.1999.02964.x}, \href
  {http://adsabs.harvard.edu/abs/1999MNRAS.310..645W} {310, 645}

\bibitem[\protect\citeauthoryear{{Xue} et~al.,}{{Xue} et~al.}{2008}]{Xue2008}
{Xue} X.~X.,  et~al., 2008, \mn@doi [\apj] {10.1086/589500}, \href
  {http://adsabs.harvard.edu/abs/2008ApJ...684.1143X} {684, 1143}

\bibitem[\protect\citeauthoryear{{Yencho}, {Johnston}, {Bullock}  \&
  {Rhode}}{{Yencho} et~al.}{2006}]{Yencho2006}
{Yencho} B.~M.,  {Johnston} K.~V.,  {Bullock} J.~S.,   {Rhode} K.~L.,  2006,
  \mn@doi [\apj] {10.1086/502619}, \href
  {http://adsabs.harvard.edu/abs/2006ApJ...643..154Y} {643, 154}

\bibitem[\protect\citeauthoryear{{Zhu}, {Marinacci}, {Maji}, {Li}, {Springel}
  \& {Hernquist}}{{Zhu} et~al.}{2016}]{Zhu2016}
{Zhu} Q.,  {Marinacci} F.,  {Maji} M.,  {Li} Y.,  {Springel} V.,   {Hernquist}
  L.,  2016, \mn@doi [\mnras] {10.1093/mnras/stw374}, \href
  {http://adsabs.harvard.edu/abs/2016MNRAS.458.1559Z} {458, 1559}

\makeatother
\end{thebibliography}



\appendix
\section{Probability distributions}
\label{appendix:stats}

\change{Here we give a short summary on how to calculate the PDF of one variable that is a function of one or more variables with known PDFs. In our case, we want to know the PDF of 
$M_{200}$ given the distributions of either scaled angular momentum, scaled energy, or both scaled angular momentum and energy.}

\change{The PDF, $p(u)$, of a variable $u$ which is a function of $x$, is given by:
\begin{equation}
	p(u) = p(x) \left| \frac{\textnormal{d}x}{\textnormal{d}u} \right| 
    \label{eq:probability_u_x}\;,
\end{equation}
where $p(x)$ is the probability of variable $x$ and where the derivative corresponds to the Jacobian of the transformation. In our case, the variable $u$ corresponds to the host halo mass, $M_{200}$, while $x$ corresponds to either the scaled angular momentum, \Ltilde{}, or the scaled energy, \Etilde{}. Replacing these variables into Eq. \eqref{eq:probability_u_x}, we obtain Eqs. \eqref{eq:mass_likelihood1d} and \eqref{eq:mass_likelihood1d_L}, that is:
\begin{equation}
	p(M_{200}|\mathbf{x}^s) = \FEE{} ~ \frac{\partial\Etilde}{\partial M_{200}} \bigg|_{\Etilde=\Etilde^s}
	\label{eq:probability_Appen_E} \;,
\end{equation}
\begin{equation}
	p(M_{200}|\mathbf{x}^s) = \FLL{} ~ \frac{\partial\Ltilde}{\partial M_{200}} \bigg|_{\Ltilde=\Ltilde^s}
    \label{eq:probability_Appen_L} \;.
\end{equation}
}

\change{To constrain $M_{200}$ using both \Etilde{} and \Ltilde{} we can extend Eq. \eqref{eq:probability_u_x} to the two-dimensional case. However, doing so entails some very involved calculations. We bypassed this step by combining the two one-dimensional cases to infer the two-dimensional expression. If the \Etilde{} and \Ltilde{} variables would be independent then we could just multiply the right-hand side terms of  Eqs. \eqref{eq:probability_Appen_E} and \eqref{eq:probability_Appen_L}. However, that is not the case, so we need to take the joint probability, \FEL{}. Furthermore, we also need to obtain the correct units, which we achieve by adding an extra $M_{200}$ factor. Putting everything together, we obtain Eq. \eqref{eq:mass_likelihood2d}, that is
\begin{equation}
    p(M_{200}|\mathbf{x}^s) = \FEL{} ~ M_{200} \frac{\partial\Etilde}{\partial M_{200}}\frac{\partial\Ltilde}{\partial M_{200}} \bigg|_{\Etilde=\Etilde^s, \;\Ltilde=\Ltilde^s}
    \label{eq:probability_Appen_EL} \;.
\end{equation}
We performed extensive tests of the three likelihoods, Eqs. \eqref{eq:probability_Appen_E}-\eqref{eq:probability_Appen_EL}, to find that they give very robust estimates of the total mass of haloes.
}

\section{Mass dependence of scaled energy and angular momentum}
\label{Appendix:Scaling}

\begin{figure}
	\includegraphics[width=\columnwidth]{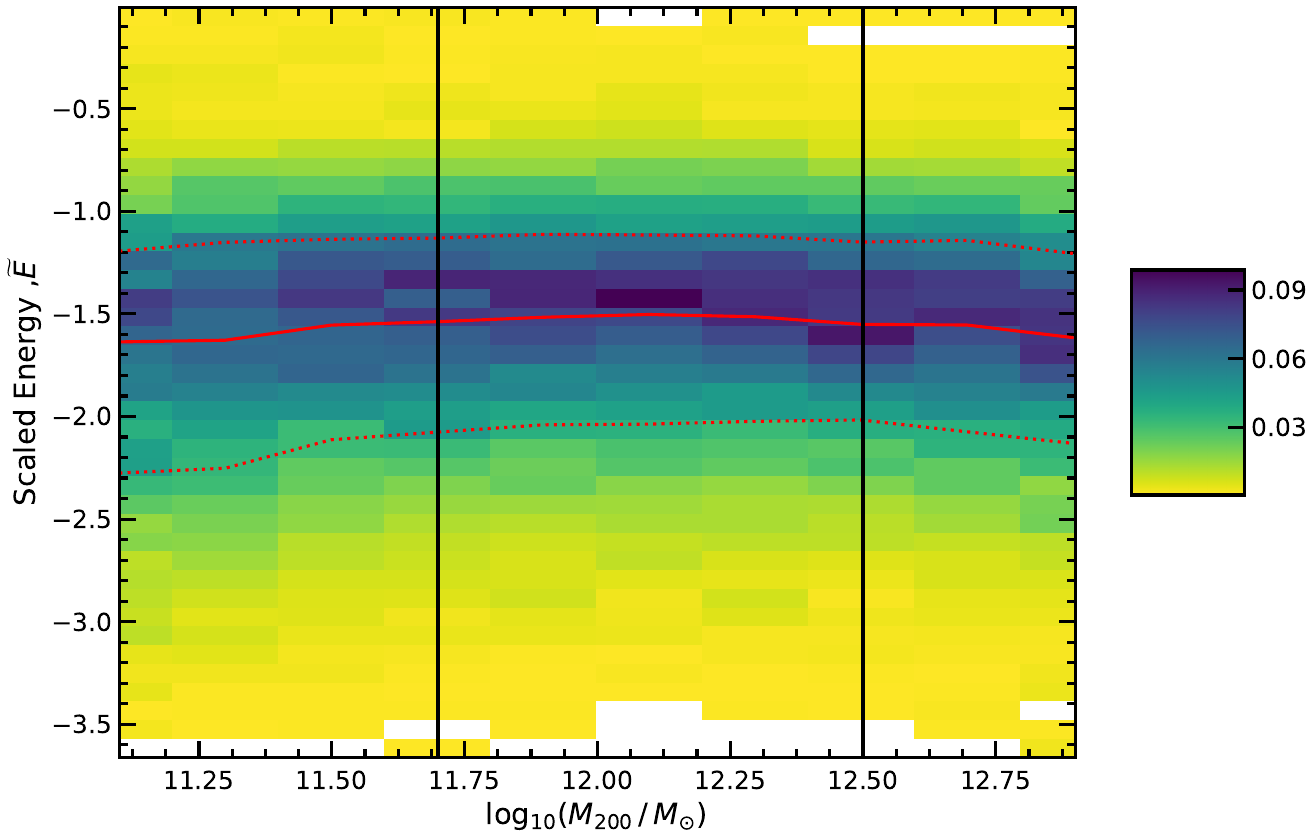}
    \\[.1cm]
    	\includegraphics[width=\columnwidth]{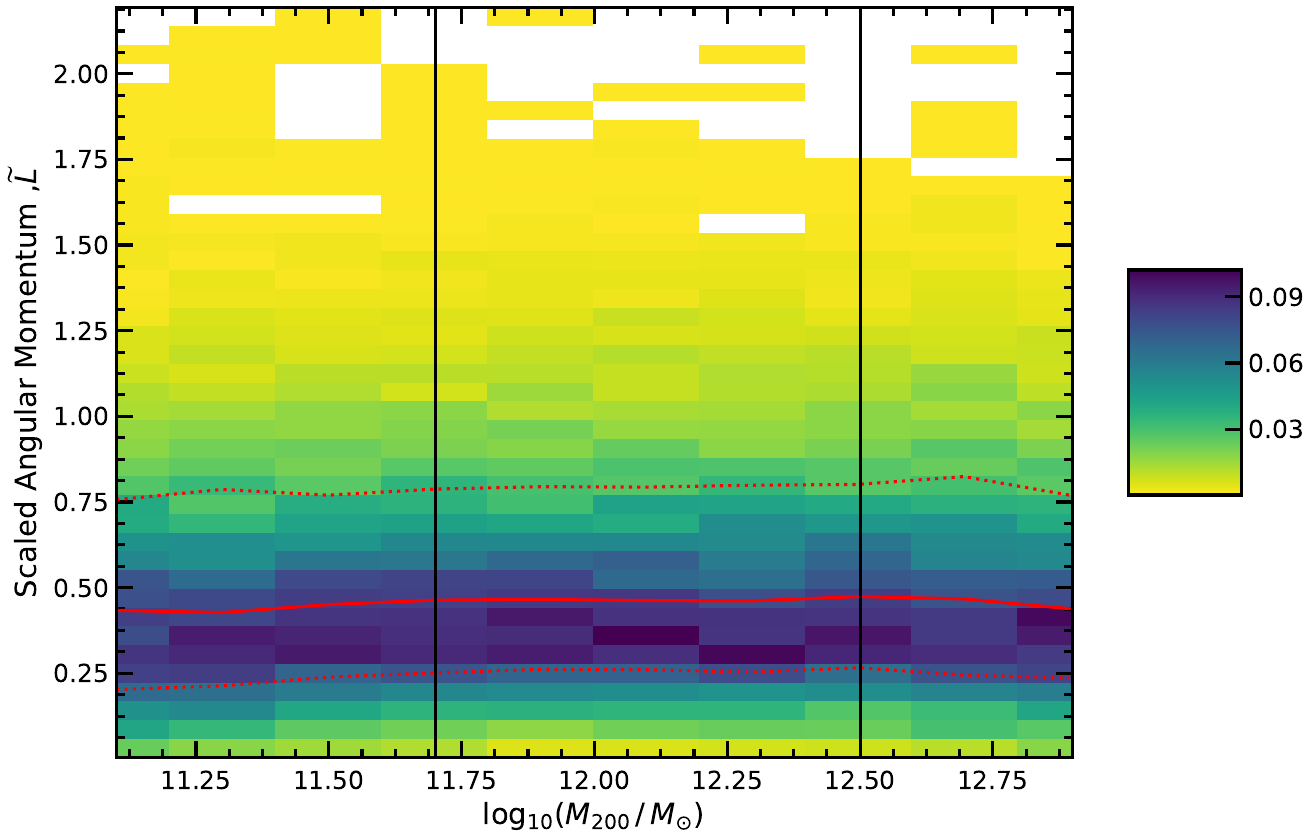}
        \smallspace{}
    \caption{ The dependence on host halo mass, $M_{200}$, of the
      scaled energy, \Etilde{} (top panel), and scaled angular
      momentum, \Ltilde{} (bottom panel), of \eagle{} satellites. The
      colour scale shows the density of points, with darker colours
      corresponding to higher density regions. The distribution is
      column normalised to each mass bin to allow easy comparison. The
      solid lines show the median values as a function of $M_{200}$,
      while the dotted lines show the 16 and 84 percentiles of the
      distribution. The two vertical lines delineate the mass range used in
      our analysis. The plots show that scaling the energy and angular
      momentum by $M_{200}^{-2/3}$ leads to quantities that are
      independent of $M_{200}$ to a very good approximation. } 
    \label{fig:appendix_scaled_E_L}
\end{figure}  

Here we test the host halo mass independence of the scaled energy and
angular momentum of satellites. We take all the luminous satellites in
the \eagle{} simulation and scale their orbital energy and orbital
angular momentum according to Eq.~\eqref{eq:sEsLEq}, that is
$\propto M_{200}^{2/3}$, where $M_{200}$ is the host mass. The
resulting distributions are shown in
Fig.~\ref{fig:appendix_scaled_E_L}.

We find that, to a very good approximation, the distributions of
\Etilde{} and \Ltilde{} are indeed the same over at least two orders
of magnitude in host mass. There are a few small departures from
universality, especially for low halo masses. This could be a
manifestation of the limited resolution of \eagle{}, which resolves
only a small fraction of the brightest satellites of $10^{11.2}\Msun$
haloes. However, this small departure from universality does not
affect our results since this work is based on hosts with masses in
the range $10^{11.7}\Msun$ to $10^{12.5}\Msun$, which corresponds to
the region between the two vertical lines in Fig. \ref{fig:appendix_scaled_E_L}.

\section{Dependence on concentration}
\label{Appendix:Concentration}

\begin{figure}
	\includegraphics[width=\columnwidth]{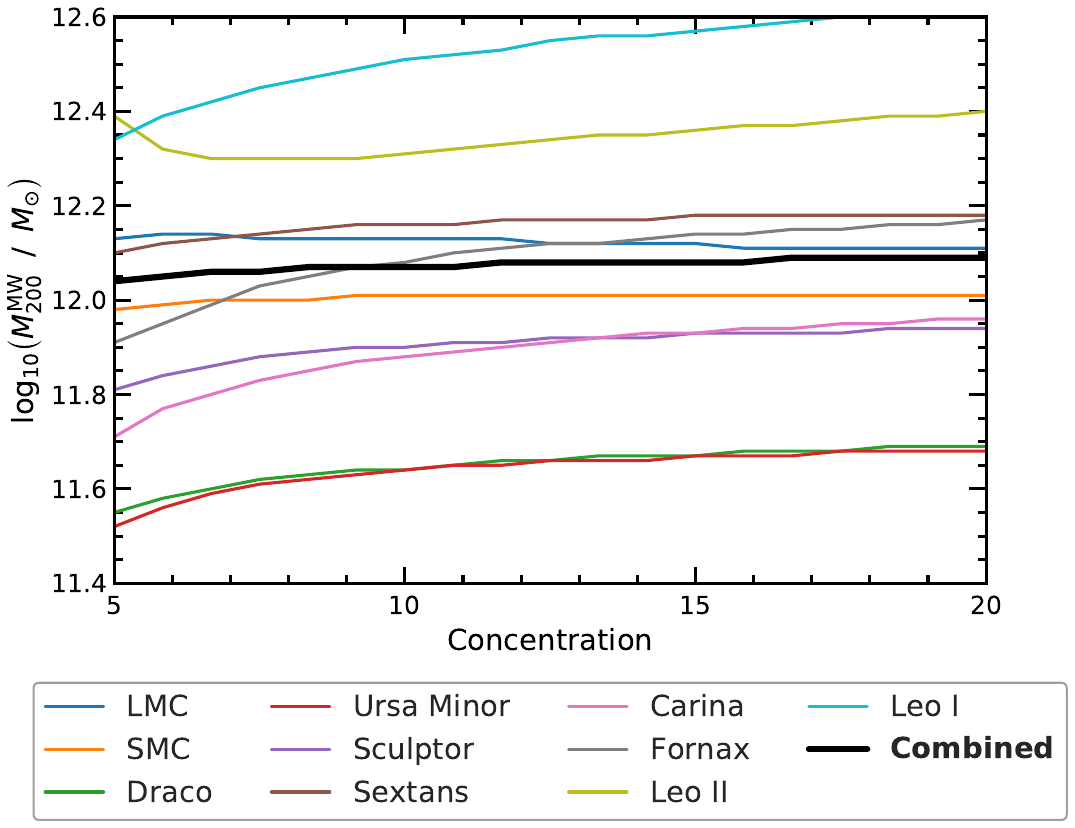}  
        \smallspace{}
        \caption{The MW total mass estimate, $M_{200}^{\mathrm{MW}}$,
          as a function of the assumed concentration of the MW halo.
          The coloured lines show the mass estimates from individual
          satellites and the black solid line shows the combined mass
          estimate. There is a very weak dependence on concentration
          --- this is especially true for the combined mass estimate,
          which remains flat over a wide range of halo concentration.}
    \label{fig:appendix_Concentration}
\end{figure}  

In Fig.~\ref{fig:appendix_Concentration} we show how the MW halo
concentration affects our mass estimate. Note that in our method
(described in Section~\ref{sec:method}) we marginalise over the
concentration parameter. The coloured lines show the mass estimates
from individual satellites and the thick black line the combined mass
estimate as a function of the assumed halo concentration. In general,
the concentration makes little difference to our estimated masses ---
this is especially true for the combined mass estimate, which remains
flat over a wide range in halo concentration. 
While not shown, we also find that the maximum likelihood values are largely independent of the assumed concentration. Thus, the 10 classical satellites studied here cannot, on their own, constrain the MW halo concentration. 
However, as we show in
Section~\ref{sec:concentration}, we can estimate the concentration of
the MW halo by combining our total halo mass estimate with
determinations of the halo mass in the inner regions of the Galaxy.

\section{Distribution of maximum likelihoods}
\label{Appendix:maximum_likelihood}

\begin{figure}
	\includegraphics[width=\columnwidth]{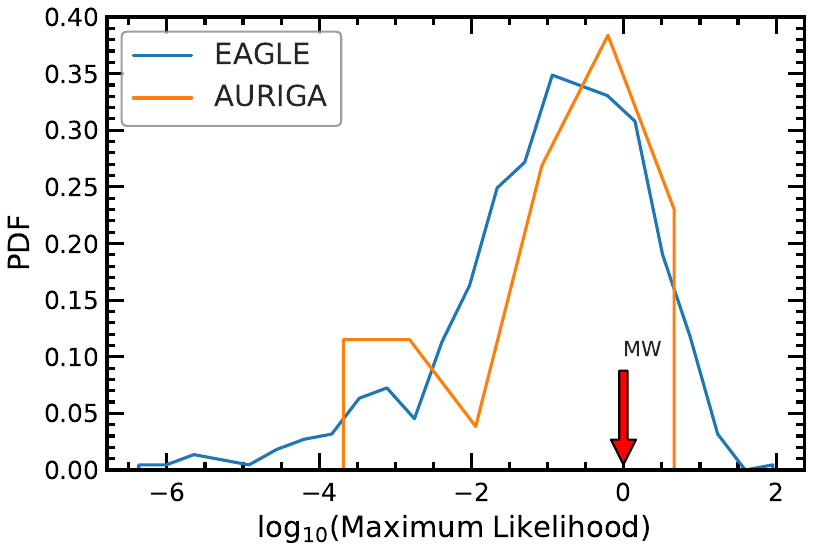}
        \smallspace{}
        \caption{ The distribution of maximum likelihood values for
          the mass determination method based on the energy and
          angular momentum of satellites. We show results for a sample
          of ${\sim}2500$ \eagle{} systems and for the 30 \auriga{}
          haloes which have a higher resolution and different galaxy
          formation models than \eagle{}. The downward pointing arrow
          shows the maximum likelihood corresponding to the MW mass
          determination, which is fully consistent with the \eagle{}
          and \auriga{} distributions. This indicates that the MW is
          not an atypical system in terms of its satellites' energy
          and angular momentum, and thus we can trust our MW
          mass determination.  }
    \label{fig:appendix_maximum_likelihood}
\end{figure}  

The MW classical satellites have at least two atypical properties:
(i)~they are distributed on a thin plane with many of the satellites
rotating within this plane, and (ii)~the satellites have a very low
velocity anisotropy indicative of circularly biased orbits. These two
characteristics place the MW satellite system in the $5\%$ and $2\%$
tails of the $\Lambda$CDM predictions
\citep{Cautun2015,Cautun2017}. This raises the concern that the
satellites may also be atypical in terms of their energy or angular
momentum distributions. If so, this could lead to biases or
untrustworthy MW mass estimates using our method.

A straightforward way to test for this is to compare the maximum
likelihood value for the MW with the corresponding values for a large
sample of $\Lambda$CDM haloes. This is shown in
Fig.~\ref{fig:appendix_maximum_likelihood}, where we plot the
distribution of maximum likelihood values for the \eagle{} and
\auriga{} mock satellite systems. We find very good agreement between
the \eagle{} and \auriga{} mocks and, more importantly, the value for
the MW lies in the central region of the $\Lambda$CDM
expectation. This indicates that we can find a range of $M_{200}$
values for the Galactic halo for which the classical satellites have
energy and angular momentum values that are fully consistent with the 
$\Lambda$CDM predictions.

\bsp	
\label{lastpage}
\end{document}